\documentclass{aa}  

\usepackage{bm}
\usepackage[citecolor=blue, linkcolor=blue, urlcolor = black, colorlinks = true]{hyperref}
\usepackage{graphicx}
\usepackage{txfonts}
\usepackage{lipsum}
\usepackage{subcaption}         
\usepackage{lscape}             
\usepackage{placeins}           
\usepackage{orcidlink}          

\usepackage{breqn} 
\everymath{\displaystyle}

\def\v{{\boldsymbol{v}}}
\def\u{{\boldsymbol{u}}}
\def\uosc{{\boldsymbol{u}_{\rm osc}}}
\def\ut{{\boldsymbol{U_{\rm t}}}}

\def \Om{{\boldsymbol{\Omega}}}

\def\et{{\boldsymbol{e_{\theta}}}}
\def \er{{\boldsymbol{e_{r}}}}
\def\ep{{\boldsymbol{e_{\varphi}}}}

\def \di{\boldsymbol{\nabla} \cdot}
\def \nab{\boldsymbol{\nabla}}

\newcommand{\adv}[2]{\ensuremath \left(\boldsymbol{#1}\cdot \boldsymbol{\nabla}\right) #2}

\newcommand{\Dd}[2]{\ensuremath\frac{\partial #1}{\partial#2}}
\newcommand{\Dt}[1]{\ensuremath\frac{\partial #1}{\partial t}}

\newcommand{\vc}[1]{\ensuremath\boldsymbol{#1}}

\begin{document}
\nolinenumbers

   \title{The impact of rotation on the stochastic excitation of \\ stellar acoustic modes in solar-like pulsators}
\subtitle{}
 \author{L. Bessila \orcidlink{0009-0007-8721-7657} \inst{1}
   \and A. Deckx van Ruys\inst{1}\fnmsep\inst{2}
   \and V. Buriasco \inst{1}
          \and
          S. Mathis\inst{1}
          \and
          L. Bugnet \orcidlink{0000-0003-0142-4000} \inst{3}
          \and
          R.A. García \orcidlink{0000-0002-8854-3776} \inst{1}
          \and
          S. Mathur \orcidlink{0000-0002-0129-0316} \inst{4,5}
          }

   \institute{Université Paris-Saclay, Université Paris Cité, CEA, CNRS, AIM, Gif-sur-Yvette, F-91191, France
   \and 
   Ecole polytechnique, Institut   Polytechnique de Paris, Palaiseau, France, 
         \and
            Institute of Science and Technology Austria (IST Austria), Am Campus 1, Klosterneuburg, Austria
        \and
            Instituto de Astrofísica de Canarias (IAC), E-38205, La Laguna, Tenerife, Spain
        \and
            Universidad de La Laguna (ULL), Departamento de Astrof\'isica, E-38206 La Laguna, Tenerife, Spain
            }

   \date{Received 12 September 2024 / Accepted 12 February 2025}

  \abstract
   {Recent observational results from asteroseismic studies show that an important fraction of solar-like stars do not present detectable stochastically excited acoustic oscillations. This non-detectability seems to correlate with a high rotation rate in the convective envelope and a high surface magnetic activity. At the same time, the properties of stellar convection are affected by rotation and magnetism.}
   {We investigate the role of rotation in the excitation of acoustic modes in the convective envelope of solar-like stars, to evaluate its impact on the energy injected in the oscillations.}
   {We derived theoretical prescriptions for the excitation of acoustic waves in the convective envelope of rotating solar-like stars. We adopted the rotating mixing-length Theory to model the influence of rotation on convection.  We used the MESA stellar evolution code and the GYRE stellar oscillation code to estimate the power injected in the oscillations from our theoretical prescriptions.}
   {We demonstrate that the power injected in the acoustic modes is insensitive to rotation if a Gaussian time-correlation function is assumed, while it can decrease by up to 60 \% for a Lorentzian time-correlation function, for a $20 \Omega_{\odot}$ rotation rate. We show that the modification of the excitation rate by rotation depends not only on the rotation rate but also on the radial and angular orders of the considered oscillation mode. This result can allow for better constraints on the properties of stellar convection by studying observationally acoustic mode excitation.}
   {These results demonstrate how important it is to take into account the modification of stellar convection by rotation when evaluating the amplitude of the stellar oscillations it stochastically excites. They open the path for understanding the large variety of observed acoustic-mode amplitudes at the surface of solar-like stars as a function of surface rotation rates.}


\maketitle

\section{Introduction}
  Acoustic waves are stochastically excited in the convective envelope of solar-like stars, mainly due to turbulent Reynolds stresses \citep[e.g.][]{goldreich_solar_1977, samadi_excitation_2001}. Standing acoustic waves, hereafter denoted as '${\rm p}$ modes', form in resonant cavities extending from the surface to an internal turning sphere, whose radial location depends on the frequency of the mode and its latitudinal degree. As pressure fluctuations induce temperature and luminosity variations, one can observe stochastically excited acoustic oscillation modes at the surface of main-sequence and evolved solar-like pulsators \citep[e.g.][]{kjeldsen_amplitudes_1995, appourchaux_corot_2008, bedding_observations_2007, michel_corot_2008, barban_solar-like_2009, ballot_accurate_2011, campante_asteroseismic_2016, mathur_detections_2022, hatt_catalogue_2023, gonzalez-cuesta_multi-campaign_2023}. Their detection has been made possible with photometric data obtained by space missions such as Convection, Rotation, and Transits \citep[CoRoT;][]{baglin_corot_2006}, Kepler/K2 \citep{borucki_finding_2008, borucki_kepler_2010, howell_k2_2014}, or Transiting Exoplanet Survey Satellite \citep[TESS; ][]{ricker_transiting_2014}. 
  The study of stellar acoustic modes provides valuable information on stellar interiors \citep[see for instance the reviews on helioseismology and asteroseismology of][]{aerts_periodic_2010, garcia_asteroseismology_2019, christensen-dalsgaard_solar_2021}, as they propagate inside stars and are affected by internal processes. For instance, acoustic modes observed at the surface of the Sun allowed for measurement of the location of the base of the convective envelope \citep[e.g.][]{christensen-dalsgaard_depth_1991, basu_solar_1997}, of the internal rotation profile \citep[from the surface down to about 0.25\, R$_\odot$;][]{schou_helioseismic_1998, thompson_internal_2003, garcia_tracking_2007, mathur_sensitivity_2008}, and sound speed and density profiles \citep{antia_nonasymptotic_1994, couvidat_rotation_2003}. Asteroseismology of solar-like stars allows one to constrain their evolutionary state \citep{garcia_can_2008, bedding_multi-site_2010, mosser_probing_2012, stello_oscillating_2015}, the mean rotation in the convective zone  \citep[e.g.][]{davies_asteroseismology_2016, benomar_asteroseismic_2018, hall_weakened_2021}, and their magnetic activity \citep[e.g.][]{lanza_stellar_2009, garcia_corot_2010}.
  
\cite{hon_search_2019} detected solar-like oscillations in about $92\%$ of red giants observed by the \textsl{Kepler} mission by using an automated code based on machine learning. On the main sequence, the detection rate of solar-like oscillations is reduced to $40\%$ \citep[e.g.][]{chaplin_evidence_2011}. In the Sun, acoustic-mode amplitudes are known to be sensitive to variations in the magnetic activity level \citep[e.g.][]{woodard_change_1985, elsworth_evidence_1990, jimenez-reyes_excitation_2003, howe_validation_2015, garcia_seismic_2024}, namely, acoustic mode amplitudes decrease with increasing activity level \citep[e.g.][]{chaplin_variations_2000, komm_solar-cycle_2000, kiefer_effect_2018}. Such activity-related variations have also been detected in other solar-like stars  \citep{garcia_corot_2010, kiefer_direct_2017,salabert_frequency_2018, santos_signatures_2018}.

\begin{figure}
     \centering
     \includegraphics[width = 8.5cm]{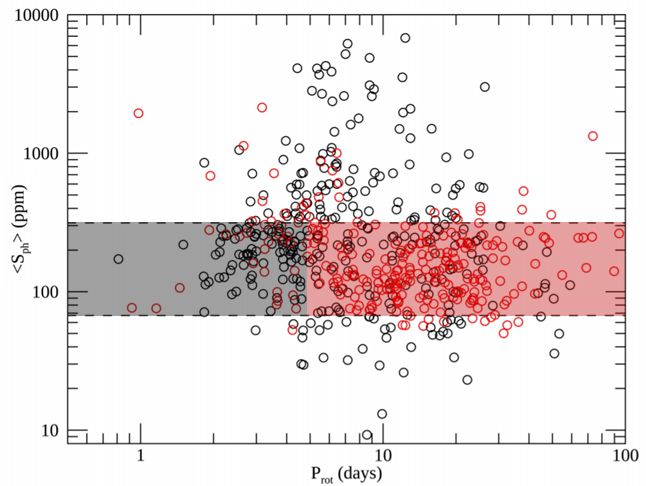}
     \caption{$S_{\rm ph}$ as a function of the surface rotation period, $P_{\rm rot}$ for $867$ main-sequence stars observed in short cadence by the \textsl{Kepler} mission. Red circles represent stars with detected oscillations, and black circles are stars without their detection. Dashed lines represent the minimum (bottom) and maximum (top) $S_{\rm ph}$ corresponding to the Sun magnetic activity. Coloured rectangles delimit the sample at a rotation period of about 5 days: in the grey area most stars do not present detectable oscillations, while in the slow-rotation regime (red area) most stars present detectable oscillations. Adapted from \cite{mathur_revisiting_2019}.}
     \label{fig:mathur}
     \end{figure}

In this framework, \cite{mathur_revisiting_2019} studied a sample of $867$ solar-like oscillating stars observed during the \textsl{Kepler} mission. According to the current theory of acoustic excitation by turbulent convection, all of these stars should present solar-like oscillations because of their convective envelope. However, asteroseismic analysis has shown that ${\rm p}$ modes are not detected in about $54\%$ of these stars (see black circles on Fig. \ref{fig:mathur}). This could result from the impact of a high level of surface magnetic activity \citep[measured by the ${\rm S}_{\rm ph}$, a photometric magnetic activity proxy; see e.g.][]{mathur_magnetic_2014, salabert_seismic_2015} on the excitation of acoustic modes. Other parameters such as rapid rotation, metallicity, or binarity could also be responsible for this non-detection. Figure \ref{fig:mathur} represents the sample of solar-like stars from the study of \cite{mathur_revisiting_2019} in the ${\rm S}_{\rm ph}$ versus surface rotation period (${\rm P}_{\rm rot}$) diagram. If we focus our attention on stars with similar magnetic activity compared to the Sun (stars located between the dashed black lines on Fig. \ref{fig:mathur}), we observe that stars rotating slower than 5 days (i.e. ${\rm P}_{\rm rot}>5$ days roughly) tend to have solar-like oscillation detection. On the contrary, solar-like oscillations seem less likely to be detected for stars rotating faster than 5 days. This indicates that not only magnetic activity but also rotation might impact the energy injected in acoustic oscillations in solar-like stars' convective envelopes. 
\par The power contained in acoustic oscillations results from a balance between the energy injection rates by turbulent Reynolds stresses \citep[e.g.][]{samadi_excitation_2001, samadi_excitation_2005} and the damping \citep[e.g.][]{grigahcene_convection-pulsation_2005}. The excitation of acoustic modes is known to take place in the external third of the star radius, where the stellar spectrum of turbulent energy was initially modelled by a Kolmogorov \citep{kolmogorov_dissipation_1941} turbulent energy spectrum \citep{goldreich_solar_1977,goldreich_wave_1990,balmforth_solar_1992}. However, according to observations, this choice for the spectrum does not represent the complete turbulent solar spectrum \citep{nesis_dynamics_1993}. In \cite{samadi_excitation_2001}, the formalism describing mode excitation was extended to all kinds of turbulent spectra. Then, \cite{belkacem_stochastic_2008} adapted the formalism to take all the components of the Lagrangian displacement of stellar non-radial oscillations into account. Later, \cite{belkacem_mode_2009} took into account the effects of slow uniform rotation on the excitation of ${\rm p}$ modes. They showed that solar-like rotation modifies acoustic mode excitation rates because of their perturbation by the Coriolis acceleration.  \cite{neiner_astronomy_2020} finally extended the formalism to rapidly rotating stars. However, none of the works described above include the direct impact of rotation on turbulent convection.
  
 In this paper, we aim to understand how p-mode amplitudes are affected by the rotation of the star, through the modification it induces on turbulent convection that excites pressure waves. Indeed, we know from fundamental fluid mechanics studies \citep[e.g.][]{chandrasekhar_hydrodynamic_1961, julien_statistical_2012} and dedicated studies as a function of stellar types \citep[e.g.][]{brun_differential_2017, brun_magnetism_2017, brun_powering_2022} that turbulent convection is strongly modified by rotation. Here, as a first step, we focus on the theoretical study of the impact of rotation on the excitation of acoustic modes and do not study the effect of rotation on the damping of these modes. We do not take into account magnetic effects; magnetic effects on the excitation and damping of the modes in rotating and magnetised stars are out of the scope of this first article and will be investigated in follow-up studies. 
\par Within this context, we used the mixing-length theory $\,$ (hereafter MLT) to assess the heat transport by convection \citep[e.g.][]{bohm-vitense_uber_1958, gough_mixing-length_1977}. \cite{stevenson_turbulent_1979} introduced the rotating-mixing length theory $\,$  (hereafter R-MLT), which describes the modification of the root-mean-square (r.m.s. velocity) and scale of turbulent convection by rotation for the mode that maximises the heat transport. \cite{barker_theory_2014} and \cite{currie_convection_2020} validated the theory and tested it with high-resolution three-dimensional non-linear hydrodynamical simulations of convection in a Cartesian box assuming the Boussinesq approximation. \cite{augustson_model_2019} generalised this theory by taking into account heat and viscous diffusions and deriving a prescription for the convective overshoot, which has been successfully compared to 3D non-linear spherical numerical simulations \citep{korre_dynamics_2021}. These prescriptions also provide interesting results to understand light-elements mixing in late-type stars \citep{dumont_lithium_2021} and convective core boundary properties in early-type stars \citep{michielsen_probing_2019}. Their model was then used to quantify the efficiency of the stochastic excitation of gravito-inertial modes, which are gravity modes influenced by rotation \citep{augustson_model_2020}. However, the impact of rotating convection on the stochastic excitation of acoustic modes has never been studied before.\\

In Section \ref{sec:model}, we present the theoretical model of acoustic mode excitation in a rotating convective envelope. We then introduce the R-MLT, presented in Section \ref{sec:rotating_convection}, to assess how rotation affects the stochastic excitation model. Finally, we implement the obtained theoretical prescription to estimate the energy injected into acoustic oscillations in typical rotating Sun-like stars. We model such stars using the Modules for Experiments in Stellar Astrophysics \citep[MESA;][]{paxton_modules_2011, paxton_modules_2013, paxton_modules_2015, paxton_modules_2018, paxton_modules_2019, jermyn_modules_2023} in order to obtain their structure and evolution, along with GYRE \citep{townsend_gyre_2013, townsend_gyre_2014} to constrain the computation of their oscillations in Section \ref{sec:numerical_estimates}. We present the main results of our work, and we conclude on its impact on the understanding of observed acoustic-mode amplitudes in rotating solar-like stars in Section \ref{sec:conclusion}.

\section{Turbulent stochastic mode excitation in rotating stars}
\label{sec:model}
\subsection{The inhomogeneous wave equation}
Following the method of \cite{samadi_excitation_2001} and \cite{belkacem_mode_2009}, we derive the inhomogeneous wave equation by taking into account the Coriolis acceleration. We neglect in the present work the action of the centrifugal acceleration and magnetic fields \citep[see][for the impact of the magnetic field]{bessila_stochastic_2024}. We use the usual spherical coordinates $(r,\theta,\varphi)$ and the corresponding unit vector basis ($\er, \et, \ep$). In the inertial reference frame, the fluid velocity field 
$\v$ is expanded into two components:
\begin{equation}
    \v(r,\theta,\varphi,t ) = \u(r,\theta,\varphi,t) + r \sin{\theta}\Omega(r,\theta)\boldsymbol{e}_{\varphi}.
    \label{rotating_velocity}
\end{equation}
Here, $\u$ is the velocity field in the rotating frame: it is then associated with both turbulent convective motion and stellar oscillation velocities. We denote $\boldsymbol{w} = r \sin {\theta}\Omega(r,\theta) \boldsymbol{e}_{\varphi}$ the velocity field associated with rotation, where $\Omega(r,\theta)$ is the rotational angular frequency. Then, assuming an axisymmetric rotation where the rotation direction is along the $\theta = 0$ axis, the angular velocity vector is  $\Om(r,\theta) = \Omega(r,\theta) \cos{\theta} \er - \Omega(r,\theta)	\sin{\theta}\et$. \\
\noindent The mass conservation equation is written as:

\begin{equation}
  \Dt{\rho} + \di{(\rho \u)} = 0.
  \label{continuity}
\end{equation}

The momentum equation, taking into account the Coriolis acceleration, boils down to \citep[see e.g.][for the rotation part]{unno_non-radial_1989}:

\begin{equation}
    \begin{aligned}
        \Dt{(\rho \u)} + &\nab :\left(\rho \u \u \right) + 
        \rho \left[\Omega \Dd{\u}{\varphi} +  2 \Om \times \u + r \sin{\theta} \left( \u \cdot \nab \right) \Omega \ep \right] \\ = 
        &\rho \boldsymbol{g} - \nab P,
     \label{momentum}
    \end{aligned}
\end{equation}
    
\noindent where $\rho$ is the density, $P$ the pressure, and $\boldsymbol{g}$ the gravitational field. 

\par To study the source term due to the turbulent background, we break down each physical quantity $f$ into an equilibrium component $f_{0}$, and an Eulerian perturbation $f_1$. Using the Cowling approximation $\boldsymbol{g}_1 = 0$ \citep{cowling_non-radial_1941} and assuming that the equilibrium quantities verify the hydrostatic equilibrium equation, we obtained the perturbed form of equations (\ref{continuity}-\ref{momentum}):

\begin{equation}
\Dt{\rho_1} + \di\left[{(\rho_0 + \rho_1) \u} \right] = 0,
\label{continuity_perturbed}
\end{equation}

\begin{dmath}
    \Dt{[(\rho_0 + \rho_1) \u]} + \nab: \left[(\rho_0 + \rho_1) \u \u \right] + (\rho_0 + \rho_1) \left[\Omega \Dd{\u}{\varphi} + 2 \Om \times \u + r \sin{\theta} \left(\u \cdot \nab \right) \Omega \ep \right] = \rho_1 \boldsymbol{g}_0 - \nab P_1.
     \label{momentum_perturbed}
\end{dmath}

The perturbed equation of state, using a second-order expansion, is written as: 
\begin{equation}
    P_1 = c_s^2 \rho_1 + \alpha_s s_1 + \alpha_{\rho_p \rho}\rho_1^2 + \alpha_{ss}s_1^2 + \alpha_{\rho s} \rho_1 s_1, 
    \label{state_equation}
\end{equation}
where: 
$$
\begin{aligned}
\alpha_s & =\left(\frac{\partial P}{\partial s}\right)_\rho, \alpha_{\rho s}=\left(\frac{\partial^2 P}{\partial \rho \partial s}\right),  \\
\alpha_{s s} & =\left(\frac{\partial^2 P}{\partial s^2}\right)_\rho, \alpha_{\rho \rho}=\left(\frac{\partial^2 P}{\partial \rho^2}\right)_s,  \\
c_s^2 & =\Gamma_1 P_0 / \rho_0.
\end{aligned}
$$
We have introduced $c_s$, the sound waves speed, and $\Gamma_1= \left(\frac{\partial \ln P }{\partial \ln \rho}\right)_s$, the first adiabatic exponent, where $S$ denotes the macroscopic entropy. We assumed that the oscillations follow an adiabatic evolution so that the Lagrangian entropy fluctuation is exclusively due to turbulence: $s_1 = s_t$. We denote $\delta s_t$ the Lagrangian turbulent entropy fluctuations and $s_t$ the turbulent Eulerian fluctuations. In the presence of rotation, the link between Eulerian turbulent and Lagrangian turbulent entropy fluctuations is: 

\begin{equation}
    \frac{d \delta s_t}{dt} = \Dt{s_1} + \Omega \partial_\varphi {(s_0+s_1)} + \adv{u}{(s_0+s_1)}. 
    \label{lagrangian_eulerian_entropy}
\end{equation}

Combining the time-derivative of Eq.(\ref{state_equation}) with Eq. (\ref{lagrangian_eulerian_entropy}), one finds: 

\begin{equation}
\begin{aligned}
 \frac{\partial P_1}{\partial t} &= c_s^2 \partial_t \rho_1 + \alpha_s \frac{d \delta s_t}{dt} - \alpha_s \vc{u} \cdot \nabla s_0 - \nabla \cdot (\alpha_s s_1 \vc{u}) + s_1 \vc{u} \cdot \nabla \alpha_s \\
 &+ \alpha_s s_1 \nabla \cdot \vc{u} + 2 \alpha_{\rho \rho} \rho_1 \frac{\partial \rho_1}{\partial t} + 2 \alpha_{ss} s_1 \frac{\partial \rho_1}{\partial t} + \alpha_{\rho s} s_1 \frac{\partial \rho_1}{\partial t}\\
&+ \alpha_{\rho s} \rho_1 \frac{\partial s_1}{\partial t} - \alpha_s \Omega \frac{\partial s_0}{\partial \varphi} - \alpha_s\Omega \frac{\partial s_1}{\partial \varphi}.
\label{eq:derivee_etat_rot}
\end{aligned}
\end{equation}

To evaluate the amplitude of stellar oscillation modes excited by turbulence, we considered that the velocity field is the superposition of turbulent convective motions and stellar oscillation velocities (respectively, $\ut$ and $\uosc$): 

\begin{equation}
    \boldsymbol{u} = \ut + \uosc + r \sin \theta \Omega \boldsymbol{e}_\varphi.
\end{equation}

\par As in previous works by \cite{samadi_excitation_2001} and \cite{belkacem_mode_2009}, we assume that the turbulent convection evolves freely and is not perturbed by the oscillations. We also neglected the non-linear terms in oscillating quantities, while keeping the terms of all orders with respect to turbulent quantities. We then found the two separated continuity equations: 

\begin{equation}
    \Dt{\rho_t} = -\di \left[(\rho_0 + \rho_t) \ut \right],
    \label{continuity_turbulent}
\end{equation}

\begin{equation}
    \Dt{\rho_{\rm osc}} = -\di \left(\rho_0\ \uosc \right).
    \label{continuity_osc}
\end{equation}

\par Differentiating (\ref{momentum_perturbed}) with respect to time and making use of Eqs. (\ref{continuity_perturbed}) and (\ref{eq:derivee_etat_rot}) while neglecting the non-linear terms in oscillating quantities yields the forced oscillation equation, also called the inhomogeneous wave equation: 

\begin{equation}
\rho_0 \left(\frac{\partial^2}{\partial t^2} - \mathcal{L}\right)\uosc + \mathcal{D} = \Dt{\mathcal{S}}.
    \label{eq:inhomogeneous}
\end{equation}

One may break down this equation into several terms. $\partial^2 \uosc / \partial t^2 - \mathcal{L} \uosc$ is the operator that governs the linear dynamics of stellar oscillations under the combined action of compressibility, buoyancy, and the Coriolis acceleration. It contains all the terms that are linear in $\uosc$, and which do not contain turbulent terms. $\mathcal{D}$ is the damping term, which describes the non-adiabatic interactions between stellar oscillations and turbulence. It includes the cross-terms with oscillating quantities $\uosc$ and turbulent quantities $\ut$. We excluded in the present formulation the radiative damping. $\partial \mathcal{S}/ \partial t$ is the source term that accounts for the forcing of the modes by the turbulent background. Only the terms that contain combinations of the turbulent quantities $s_t$, $\rho_t$, $\ut$, are part of it. 
\noindent There is a strong analogy between this inhomogeneous equation and a mechanical forced mass-spring system: $\mathcal{D}$ is homologous of a mechanical damping, while $\mathcal{S}$ is an excitation source which forces the oscillations. 

\noindent We found the following expressions: 
\begin{equation}
    \begin{aligned} & \mathcal{L}=\boldsymbol{\nabla}\left[\alpha_s \uosc \cdot \boldsymbol{\nabla} s_0+c_s^2 \boldsymbol{\nabla} \cdot \left(\rho_0 \uosc\right)\right]-\boldsymbol{g} \boldsymbol{\nabla} \cdot\left(\rho_0 \uosc\right) \\ & -\rho_0 \boldsymbol{\Omega} \frac{\partial^2 \uosc}{\partial t \partial \varphi}-2 \rho_0 \boldsymbol{\Omega} \times \frac{\partial \uosc}{\partial t}-\rho_0 r \sin \theta \frac{\partial \uosc}{\partial t} \cdot \boldsymbol{\nabla} \boldsymbol{\Omega} \boldsymbol{e}_{\varphi},\end{aligned}
\end{equation}
\begin{equation}
    \begin{aligned} \mathcal{D}= & \frac{\partial}{\partial t}\left[\frac{\partial\left(\rho_{\mathrm{t}} \uosc\right)}{\partial t}+2 \boldsymbol{\nabla}:\left(\rho_0 \uosc \ut\right)+\rho_{\mathrm{t}} \boldsymbol{\Omega} \frac{\partial \uosc}{\partial \varphi}\right. \\ & +2 \rho_{\mathrm{t}} \boldsymbol{\Omega} \times \uosc+\rho_{\mathrm{t}} r \sin \theta\left(\uosc \cdot \boldsymbol{\nabla} \Omega\right) \boldsymbol{e}_\varphi \\ & \left.+\boldsymbol{\nabla}\left(\alpha_s \uosc \cdot \boldsymbol{\nabla} s_1+c_s^2 \boldsymbol{\nabla} \cdot\left(\rho_{\mathrm{t}} \uosc\right)\right)\right].\end{aligned}
\end{equation}

\subsection{Source terms }
\par We focus now on the excitation source for the oscillations.  One can break down the dominant source term $\partial \mathcal{S}/\partial t$ into several components to identify the various origins of the energy injected into the modes: 
\begin{equation}
    \mathcal{S} = \mathcal{S}_R + \mathcal{S}_S + \mathcal{S}_\Omega + \mathcal{S}_N,
\label{eq:sources}
\end{equation}
where $\mathcal{S}_R$ is the Reynolds stresses source terms expressed as: 
\begin{equation}
    \Dt{\mathcal{S}_R} = -\frac{\partial}{\partial t} \Bigg[ \nabla: \Big( \rho \ut \ut \Big)\Bigg];
\end{equation}
$\mathcal{S}_S$ is the entropy fluctuations source term,
    \begin{equation}
        \mathcal{S}_S= - \nabla \left(\alpha_s \frac{d \delta s_t}{dt} - \nabla \cdot (\alpha_s s_t \ut) + s_t \ut \cdot \nabla \alpha_s \right);
    \end{equation}
$\mathcal{S}_\Om$ is the source term due to rotation, and its gradient,
    \begin{equation}
   \frac{\partial \mathcal{S}_\Omega}{\partial t} =  - \frac{\partial}{\partial t} \left[ \rho_t \Omega \frac{\partial \ut}{\partial \varphi} + 2 \rho_t \vc{\Omega} \times \ut + \rho_t r \sin \theta (\ut \cdot \nabla) \Omega \hat{\vc{e}}_\varphi \right].
    \label{termessource}
\end{equation}

Finally, one has 
\begin{equation}
    \begin{aligned} \frac{\partial \mathcal{S}_N}{\partial t} = &\boldsymbol{\nabla}\left[c_s^2 \boldsymbol{\nabla} \cdot\left(\rho_{\mathrm{t}} \ut\right)\right]-\boldsymbol{g} \boldsymbol{\nabla} \cdot\left(\rho_{\mathrm{t}} \ut\right) \\ & -\frac{\partial^2}{\partial t^2}\left(\rho_{\mathrm{t}} \ut\right)+\mathcal{L}_{\mathrm{t}}, \end{aligned}
\end{equation}
\noindent where $\mathcal{L}_t$ contains the linear terms that do not contribute to the excitation. Previous studies \citep[e.g.][]{samadi_modeling_2008} have shown that the contribution of entropy fluctuations accounts for less than $10 \%$ of the total power injected in the modes in solar-like stars. We neglected the contribution of the entropy fluctuation source term $\mathcal{S}_S$ in this work. Moreover, \cite{belkacem_mode_2009} showed that $\mathcal{S}_{\Omega} \sim \mathcal{M} \mathcal{S}_R$ where $\mathcal{M}$ is the Mach number. The rotational contribution $\mathcal{S}_\Omega$ is then negligible compared to the Reynolds stresses term $\mathcal{S}_R$ in the subsonic regime, which is valid inside the Sun and solar-like pulsators. \\
Finally, as shown in \cite{samadi_excitation_2001}, $\partial \mathcal{S}_N/ \partial t$ does not significantly contribute to the stochastic excitation. Indeed, the terms contained are either negligible compared to the Reynolds stresses term $\mathcal{S}_R$, or linear with respect to turbulent quantities. As detailed in \cite{samadi_excitation_2001} in their section 4.1, linear terms in turbulent quantities vanish when integrated over the convective envelope to compute the resulting injected power.
From this point on, we then consider that the excitation source terms are reduced to the Reynolds stresses term $\mathcal{S}_R$ alone.  


\subsection{Mode amplitudes with rotation}
\label{sub:modes_amplitudes}

\par Using the inhomogeneous wave equation, we sought the mean square amplitude of $\uosc$, which is directly related to the power $\mathcal{P}$ injected into each mode of a given radial, latitudinal, and azimuthal order $(n,\ell,m)$. Following \cite{samadi_excitation_2001} and \cite{belkacem_mode_2009}, we obtained:

\begin{equation}
    \mathcal{P} = \eta \langle \lvert A \rvert ^2 \rangle I \omega_0^2, 
\end{equation}
where $\eta$ is the damping rate of the waves and $\omega$ is the mode frequency in the inertial reference frame. We introduce the mode inertia $I = \iiint_{\mathcal{V}} \rho_{0} d^{3} r\left(\boldsymbol{\xi}^{\star} \cdot \boldsymbol{\xi}\right)
$, where $\boldsymbol{\xi}$ is the mode displacement, $\boldsymbol{\xi}^{\star}$ denotes its complex conjugate and we perform the integral over the whole star volume $\mathcal{V}$. In the case of acoustic modes studied here, we can neglect the inertia correction introduced in the rapid rotation regime by \cite{neiner_astronomy_2020}.
\par The pulsational displacement $\delta \vc{r_{\rm osc}}$ and the velocity $\uosc$ are written using the adiabatic eigendisplacement $\vc{\xi}(\vc{r},t)$, that is with no forcing, and an instantaneous amplitude $A(t)$ due to the turbulent forcing \citep{unno_non-radial_1989}:

\begin{equation}
\delta \vc{r}_{\rm osc} = \frac{1}{2} \bigg( A(t) \vc{\xi}(\vc{r}) e^{i\omega_0 t} + \rm{c.c.} \bigg),
\end{equation}

\noindent where $\rm c.c.$ indicates the complex conjugate.
In this framework, $\vc{\xi}(\vc{r})$ is expanded on  the vectorial spherical harmonics basis \citep[e.g.][]{unno_non-radial_1989}: 

\begin{equation}\vec{\xi}(\vec{r})=\left(\xi_{r ; \ell, m,n} \vec{e}_r+\xi_{h ; \ell, m,n} \vec{\nabla}_H+\xi_{T ; \ell, m,n} \vec{\nabla}_H \times \vec{e}_r\right) Y_{\ell, m} (\theta, \varphi),
\end{equation}
where $\xi_r, \xi_H$, and $\xi_T$ are the radial, horizontal, and toroidal functions of the displacement eigenfunction, respectively. Then, the oscillation velocity $\uosc$ is given by 
\begin{equation}
\uosc=A(t)\bigg[i \widehat{\omega}_0 \boldsymbol{\xi}(\vc{r})-(\boldsymbol{\xi}(\vc{r}) \cdot \boldsymbol{\nabla} \Omega) r \sin \theta \boldsymbol{e}_\varphi\bigg] \mathrm{e}^{\mathrm{i} \omega t},
\end{equation}
where $\widehat{\omega}_0 =\omega_0+m \Omega$, and $\omega_0$ is the wave frequency in an inertial reference frame. In the latter expression, the temporal variation of the instantaneous amplitude $A(t)$ was neglected, as the oscillation periods are much shorter than their lifetime: $d \ln A(t)/dt \ll \omega_0$ \citep[see e.g.][]{samadi_excitation_2001}. 
In the framework of this paper, we restrict ourselves to uniform rotation as a first step, such that $\uosc$ becomes 

\begin{equation}
\uosc=A(t) i \widehat{\omega}_0 \boldsymbol{\xi}(\vc{r},t)) \mathrm{e}^{\mathrm{i} \omega_0 t}.
\label{eq:amplitudeosc}
\end{equation}

\begin{figure}[!ht]
    \centering
    \includegraphics[width=\linewidth]{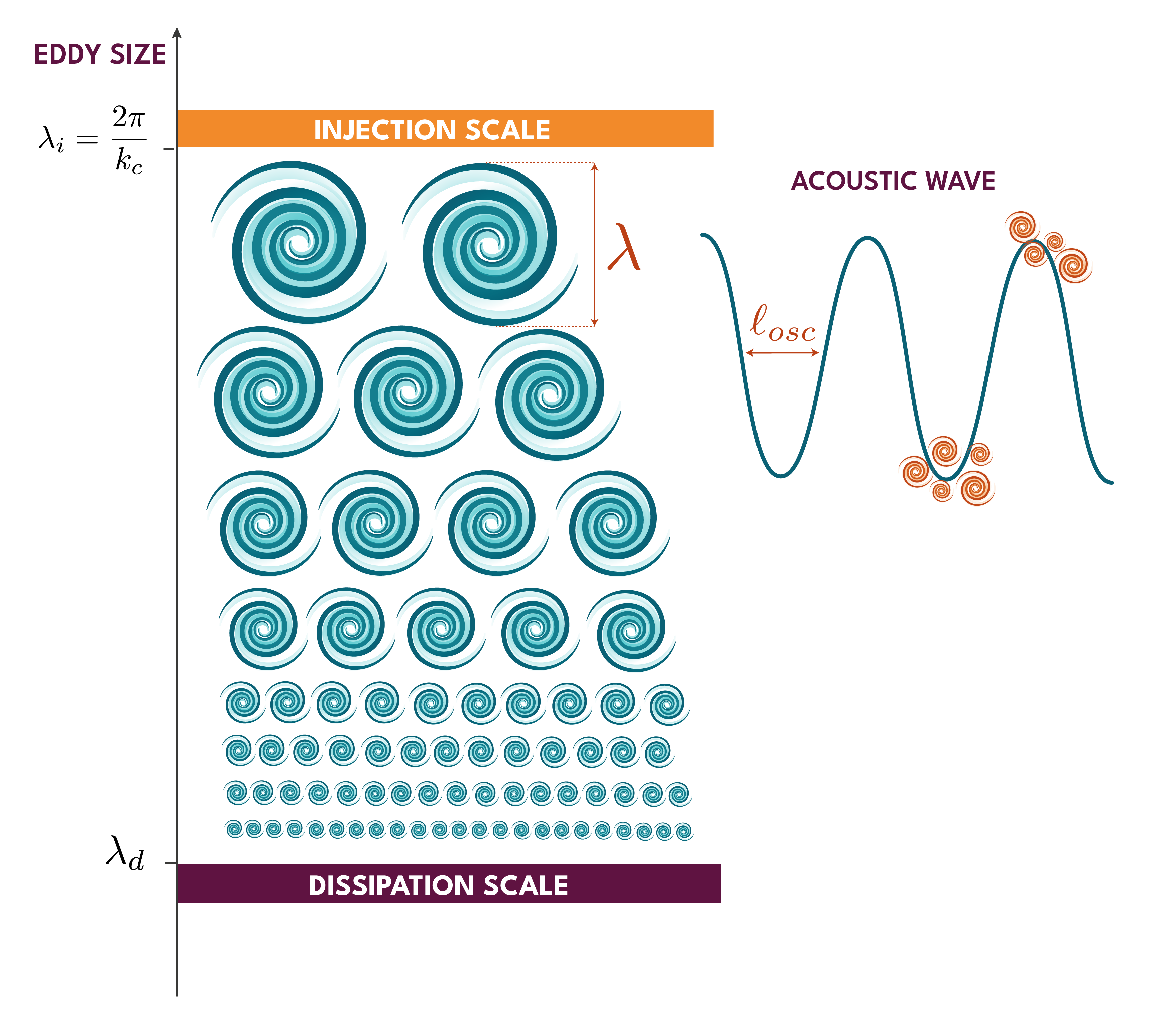}
    \caption{Illustration of the scale separation assumption for the stochastic excitation. In the turbulent cascade, $\lambda_i$ denotes the energy injection scale in the framework of the turbulent cascade, while $\lambda_d$ is the dissipation scale. The size of the largest eddy in the inertial range is estimated by the mixing length $\lambda_i$. The size of an arbitrary eddy is denoted $\lambda$. All the eddies in the cascade contribute to the stochastic excitation.}
    \label{fig:scale-sep}
\end{figure}

Moreover, for acoustic waves studied here $\omega_0 \gg \Omega$. We then assume that $\widehat{\omega}_0 \approx \omega_0$. The mean squared amplitude is then \citep[e.g.][]{samadi_excitation_2001, belkacem_mode_2009}:

\begin{equation}
\left\langle|A|^2\right\rangle=\frac{1}{8 \eta\left(\omega_0 I\right)^2}C_R, 
\end{equation}
where $C_R$ is the Reynolds-stress contribution: 
\begin{equation}
C_R=\int d^3 x_0 \int_{-\infty}^{+\infty} d^3 r d \tau e^{-i \omega_0 \tau}\Bigg\langle(\vec{\xi} \cdot \mathcal{S_R})_1(\vec{\xi} \cdot \mathcal{S_R})_2\Bigg\rangle. 
\end{equation}
Here the operator $\langle (.)(.)\rangle$ denotes the statistical average performed on an infinite number of independent realisations. Subscripts 1 and 2 represent the values taken at spatiotemporal positions $(\boldsymbol{x}_0 - \boldsymbol{r}/2, -\tau/2)$ and $(\boldsymbol{x}_0 + \boldsymbol{r}/2, \tau/2)$, respectively.
To compute the contribution $C_R$, we follow the  formalism from \cite{samadi_excitation_2001} which is built on three main assumptions: 

\begin{enumerate}
    \item  It is assumed that the wavelength of the eigenfunctions is large compared to the typical lengthscale of turbulence: this assumption is called the separation of scales. To apply this formalism, one must then compare the oscillation characteristic lengthscale $\ell_{\rm osc}$ to the turbulent characteristic length of the eddies $\lambda$. Only the turbulent eddies such that $\lambda \lesssim \ell_{\rm osc}$ contribute to the stochastic excitation (see illustration in Figure \ref{fig:scale-sep}, and \cite{samadi_excitation_2001} for more details). Using the softwares MESA and GYRE, we find that this scale-separation assumption is valid for acoustic modes in solar-like stars. Indeed, the injection scale verifies $\lambda_i \lesssim \ell_{\rm osc}$.
    \item The turbulent background is considered homogeneous: thermodynamical quantities and the oscillations vary on scale lengths much larger than the contributive turbulent eddies \citep[see e.g.][]{goldreich_wave_1990, samadi_excitation_2001}. 
    \item Finally, isotropic turbulence is assumed.
    Although rotating turbulence is anisotropic \citep[e.g.][]{aurnou_rotating_2015}, this simplification is necessary to combine the stochastic excitation formalism by \cite{samadi_excitation_2001} and Rotating MLT, which reproduces well the properties of the convective mode that transports the most heat in rotating anisotropic convective flows \citep[e.g.][]{barker_theory_2014, currie_convection_2020, vasil_rotation_2021}. Tackling anisotropic turbulence with rotation is subject to forthcoming developments. 
\end{enumerate}

\begin{figure}[!ht]
    \centering
\includegraphics[width=\linewidth]{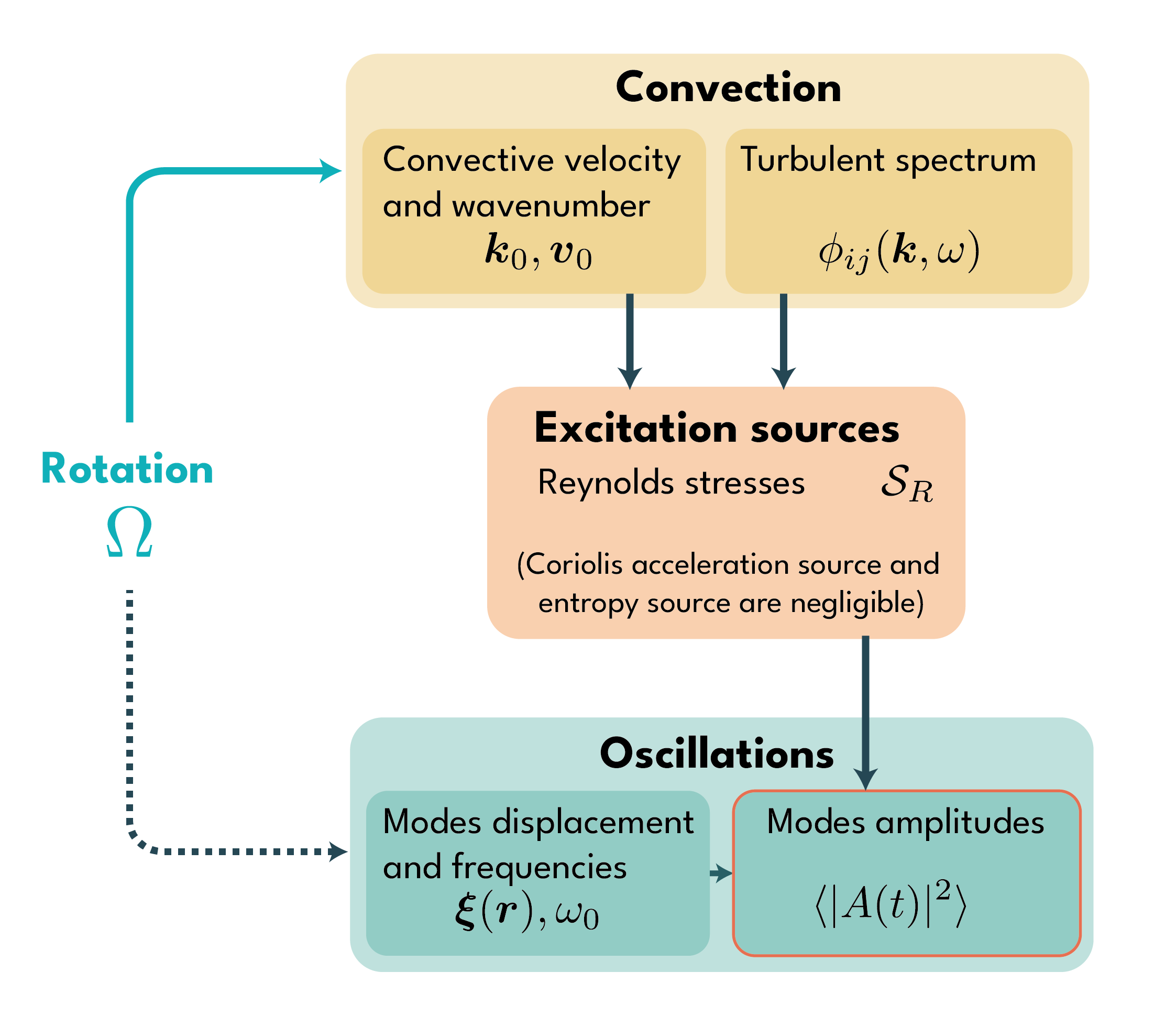}
    \caption{Interdependencies in the stochastic excitation mechanism taking rotation into account. The light blue arrow denotes the direct impact of rotation on convection we take into account, while dotted arrows represent phenomena we neglect. Dark arrows represent the general interdependencies in the formalism describing the turbulent stochastic excitation of stellar oscillations.}
    \label{fig:interdependencies}
\end{figure}

Moreover, as underlined in \citep{belkacem_waves_2008}, modes with frequencies between 1 mHz and 5 mHz and low angular degree $\ell$ are essentially radial that is $\xi_h \ll \xi_r$.  This is the case for the solar p-modes considered in the present study. We also use the plane-parallel approximation for acoustic modes:  it is justified in the present case by the fact that excitation occurs in the uppermost part of the convection zone $(r / R \approx 1)$. Such an approximation is valid when the condition $r k_{\mathrm{\rm osc}} \gg 1$ is fulfilled in the excitation region, $k_{\text {osc }}$ being the local wavenumber \cite[e.g.][]{belkacem_stochastic_2008}. In this framework, one can write the Reynolds stresses contribution as in \cite{samadi_excitation_2001}:

\begin{equation}
    C_R=\frac{16}{15} \pi^3 \int_{\mathcal{V_{CZ}}} d^3 x_0 \rho_0^2\left|\frac{\mathrm{d} \xi_r} {\mathrm{~d} r}\right|^2 Y_{\ell,m} Y^{*}_{\ell,m}\hat{S}_R\left(r,\theta,\omega_0\right), 
    \label{eq:reynolds-int}
\end{equation}
where we have introduced the spherical harmonics $Y_{\ell,m}$ and their complex conjugates $Y_{\ell,m}^{*}$. The integral is performed over the whole convective zone volume $\mathcal{V_{CZ}}$. We also introduced the source function

\begin{equation}
\hat{S}_R\left(r,\theta,\omega_0\right)=\int \frac{d k}{k^2} E^2(k) \int d \omega \chi_k\left(\omega+\omega_0\right) \chi_k(\omega). 
\label{eq:s_hat_def}
\end{equation}
Here, $E(k)$ is the spatial kinetic energy spectrum in turbulence, and $\chi_k(\omega)$ is the associated eddy time-correlation function following \cite{stein_generation_1967}. With rotation, these kinetic energy and time-correlation spectra depend both on the radius and the colatitude of the studied location inside the convective zone. For this reason, Eq. (\ref{eq:reynolds-int}) is now a multi-dimension integral where we cannot separate the radial part from the latitudinal one. In the non-rotating case, where the source function only depends on the radius we recover the result by \cite{samadi_excitation_2001}. 

\par It must also be noted that even if rotation does not add any significant source term in equation (\ref{eq:inhomogeneous}), it indirectly influences mode excitation:
\begin{enumerate}
    \item Waves are affected by the rotation so that they become inertio-acoustic waves. This changes both the frequencies of the modes $\omega_0$, which are shifted, and the eigenfunctions of the displacement $\boldsymbol{\xi}$. For acoustic modes in main-sequence solar-like stars, this effect is treated as a first-order perturbation \citep[e.g.][]{reese_acoustic_2006} so we neglect it in the present work.
    \item Convective properties strongly affect the source terms. In the MLT theoretical framework, this happens through a change of the convective wavenumber $k_0$ and the convective r.m.s. velocity $v_0$. For such an approach, rotation tends to weaken convection \citep{stevenson_turbulent_1979, barker_theory_2014, augustson_model_2019}, for the convective mode that transports the most energy. Rotation also affects turbulent convection through a change in the kinetic energy spectrum \citep{zhou_phenomenological_1995, mininni_rotating_2010}. This effect is paramount in the present model and discussed in Section \ref{sec:rotating_convection}.
    \item The stellar structure is modified by the centrifugal acceleration. The star's thermodynamic quantities (density, pressure, entropy) profiles change as well as the stellar oscillation mode resonant cavity and as a consequence the mode eigenfrequencies \citep{espinosa_lara_self-consistent_2013, reese_oscillations_2021, mombarg_first_2023}. This is the so-called indirect effect on stellar oscillations \citep{gough_effect_1990}. We neglect this effect in the present work.
\end{enumerate}
Figure \ref{fig:interdependencies} illustrates these interdependencies.

\subsection{Turbulent spectrum}
\par One has to proceed with the derivation of $\hat{S}_R$. We used the descriptions of temporal and spatial properties of the turbulent spectra.
\subsubsection{Temporal spectrum}
Various choices for the eddy time-correlation function $\chi_k(\omega)$ exist in the literature \citep[e.g.][]{goldreich_solar_1977, belkacem_turbulent_2010}. Some theoretical formulations assume a Gaussian function \citep[see e.g.][]{goldreich_solar_1977, balmforth_solar_1992, houdek_amplitudes_1999, samadi_excitation_2001}: 
\begin{equation}
        \chi_k^G(\omega)=\frac{1}{\sqrt{\pi} \omega_k} e^{-\frac{\omega^2}{2 \omega_k^2}},
\label{eq:chi_k_gauss}
\end{equation}
where $\omega_k$ is the frequency of the turbulent eddy associated with the wavenumber $k$, defined in Eq. (\ref{eq:turnover_puls}).
Other works find that a Lorentzian gives a better agreement with solar observations and numerical simulations \citep[e.g.][]{samadi_numerical_2003, belkacem_turbulent_2010, philidet_interaction_2023}: 
\begin{equation}
    \chi_k^L(\omega)=\frac{1}{\pi \omega_k} \frac{1}{1+\left(\frac{\omega}{\omega_k}\right)^2},
\label{eq:chi_k_lorentz}
\end{equation}
\noindent where we normalise this function to 1: 
\begin{equation} 
\int_{-\infty}^{+\infty} d \omega \chi_k=1.
\end{equation}

\begin{figure*}[!h]
    \includegraphics[width=0.33\linewidth]{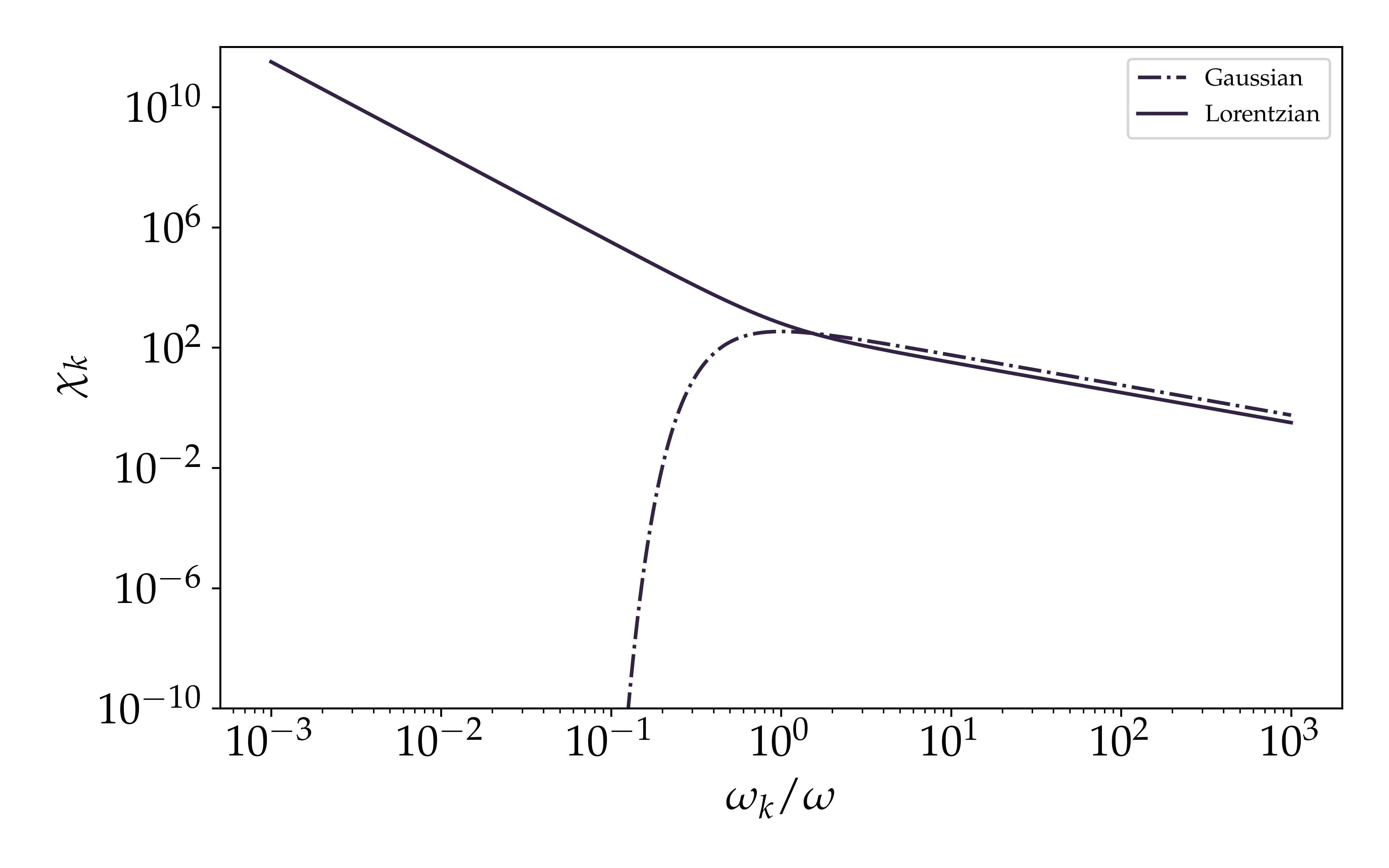}
    \includegraphics[width=0.33\linewidth]{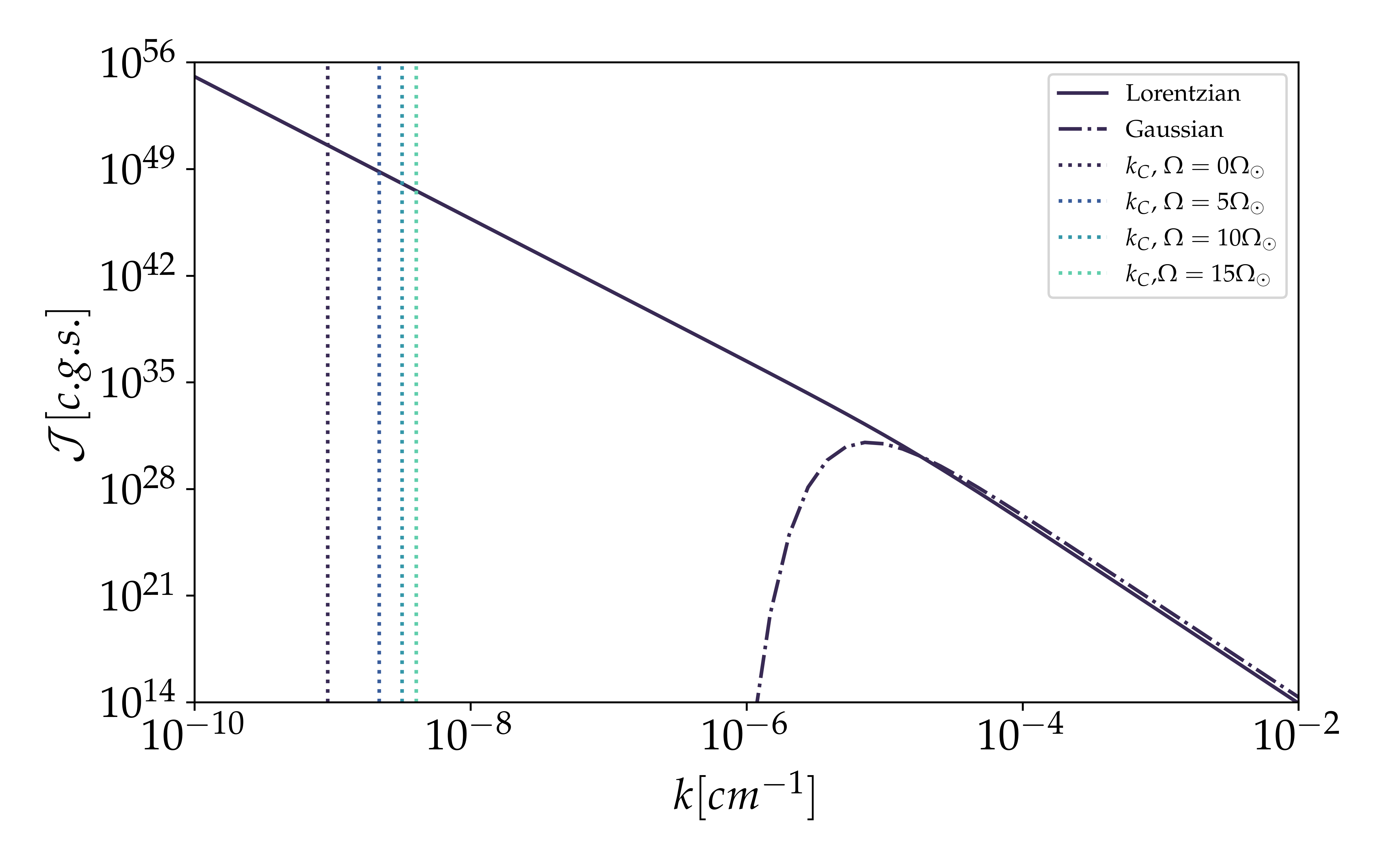}
    \includegraphics[width=0.33\linewidth]{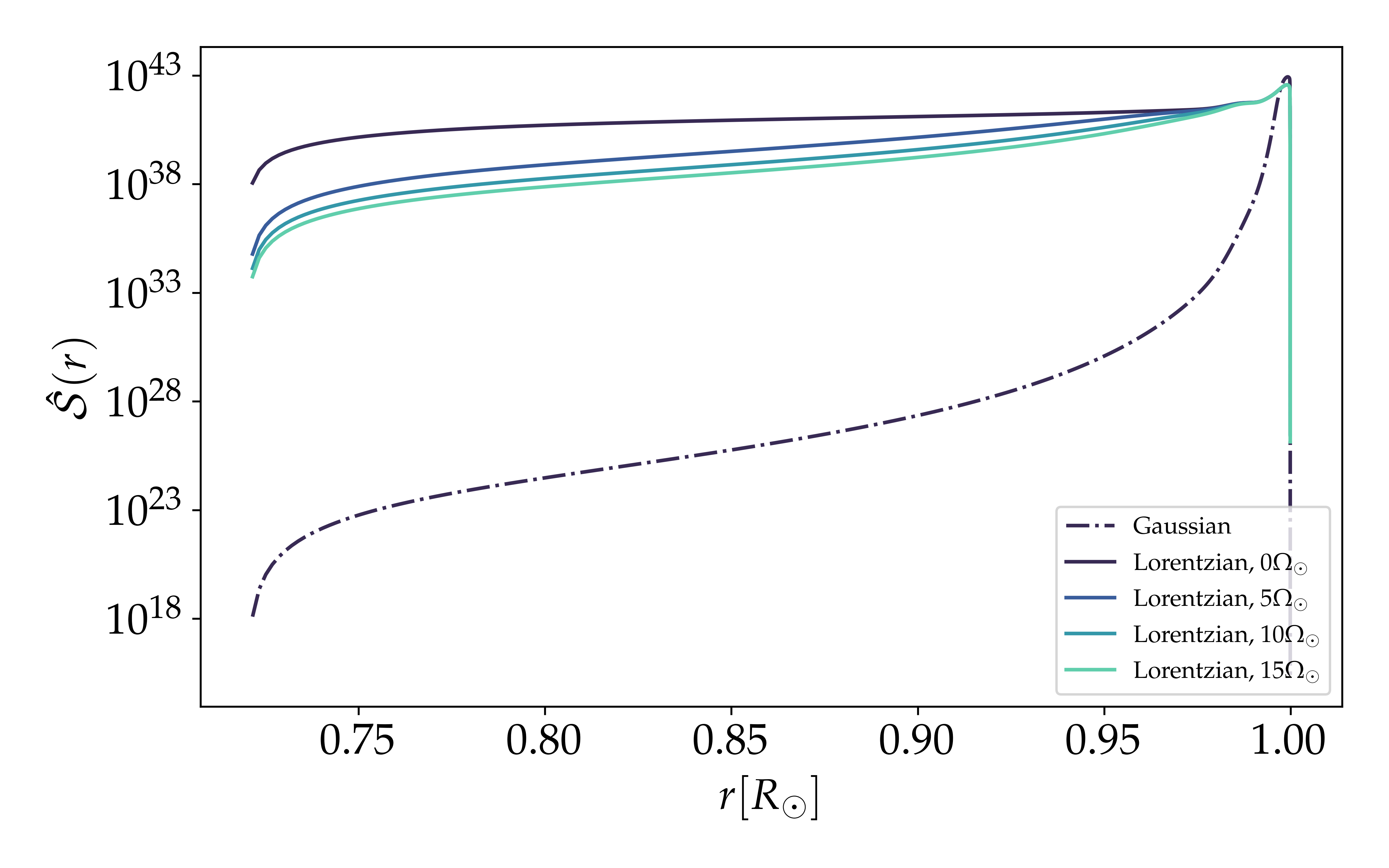}
    \caption{(Left) Comparison between a Lorentzian eddy-time correlation function (following Eq. \ref{eq:chi_k_lorentz}) and a Gaussian time-correlation function (Eq. \ref{eq:chi_k_gauss}), both with $\omega = 10^{-3} s^{-1}$. (Centre) Comparison between a Lorentzian eddy-time correlation function and a Gaussian for the integrand $\mathcal{J}(k)$ of $\hat{\mathcal{S}}$, at $r = 0.86 R_{\odot}$ and $\theta=0$. As rotation modifies the lower integration bound of $\mathcal{J}$ for low wavenumbers, it only has a significant influence on the Lorentzian eddy-time correlation. (Right) Influence of the choice for the eddy-time correlation on $\hat{\mathcal{S}}$. Rotation only affects this term when a Lorentzian is chosen.}
    \label{fig:lorentz-gauss}
\end{figure*}

 We then define the frequency component of the integral (\ref{eq:s_hat}): 
 \begin{equation}
    \mathcal{I}(\omega_0, k) \equiv \int_{-\infty}^{+\infty} \chi_k(\omega) \chi_k(\omega + \omega_0) d \omega,
    \label{eq:i_omega}
\end{equation}
\noindent where $\chi_k = \chi_k^G$ for a Gaussian time-correlation, and $\chi_k = \chi_k^L$ for a Lorentzian time-correlation function. The comparison between those functions is illustrated in Fig. \ref{fig:lorentz-gauss}.

\subsubsection{Kinetic energy spectrum}
\par The last step to compute $\hat{S}_R$ following Eq. (\ref{eq:s_hat_def}) is to proceed with the integral over the wavenumber $k$, taking into account the kinetic energy spectrum: 

\begin{equation}
\hat{S}_R\left(r, \theta, \omega_0\right)=\int_{k_c}^{+\infty} \frac{d k}{k^2} E^2(k) \mathcal{I}(\omega_0,k).
\label{eq:s_hat}
\end{equation}

\noindent The eddies turnover frequency is \citep{balmforth_solar_1992}: 
\begin{equation}
    \omega_k = k u_k,
    \label{eq:turnover_puls}
\end{equation}
where $u_k$ is the characteristic velocity of the eddy with wave number $k$. It is related to the kinetic energy spectrum $E(k)$ \citep{stein_generation_1967}:  
\begin{equation}
    u_k^2 = \int_k^{2k} E(k) dk.
    \label{eq:stein_uk}
\end{equation}
We first derive the expression for the kinetic energy spectrum. For $k > k_c$, where $k_c$ is the injection scale ($k_c$ corresponds to the largest eddy), we use a Kolmogorov energy spectrum \citep{kolmogorov_dissipation_1941}: 
\begin{equation}
    E(k) = \gamma \left(\frac{k}{k_c} \right)^{\alpha}, 
    \label{eq:energy_normalised}
\end{equation}
where $\gamma$ is a normalisation factor. The turbulent cascade conserves the overall kinetic energy: 
\begin{equation}
    \int_{k_c}^{+\infty} \gamma E(k) = \frac{\phi}{2}w^2, 
\end{equation}
where $w$ is the vertical r.m.s. velocity and $\phi$ is the anisotropy factor as defined by \cite{gough_mixing-length_1977}: 
\begin{equation}
    \phi \equiv \frac{\langle u_c^2\rangle}{\langle w_c^2 \rangle},
\end{equation}
$w_c$ being the vertical convective velocity and $u_c$ the convective velocity.
In the following, we work within the isotropic framework of the rotating mixing-length theory so that the convective velocity is the same in every direction: $\phi = 3$, and $w_c = u_c$ the r.m.s. convective velocity. The normalisation factor $\gamma$ is then: 
\begin{equation}
    \gamma = \frac{(-\alpha - 1) \phi u_c^2}{2 k_c}. 
\end{equation}
We also introduce the reduced wavenumber: 
\begin{equation}
    K \equiv \frac{k}{k_c}.
    \label{eq:def_K}
\end{equation}
Eq. (\ref{eq:stein_uk}) gives
\begin{equation}
    u_k = \sqrt{\beta} K^{\frac{1+\alpha}{2}}, 
\end{equation}
where 
\begin{equation}
    \beta = \frac{\phi}{2} u_c^2 (1 - 2^{1+\alpha}).
\end{equation}
One can deduce the turnover frequency for the eddy of wavenumber $k$, following  Eq.(\ref{eq:turnover_puls}): 
\begin{equation}
    \omega_k = \delta K^{\frac{3 + \alpha}{2}},
    \label{eq:turnover_normalised}
\end{equation}
where: 
\begin{equation}
    \delta = k_c \sqrt{\beta}.
\end{equation}
Finally, we compute the integral over the wave number $k$, assuming a Lorentzian eddy time-correlation spectrum defined by Eq. (\ref{eq:chi_k_lorentz}): 
\begin{equation}
    \hat{S}_R\left(r, \theta, \omega_0\right) = \int_{k_c}^{+\infty} \frac{E(k)^2}{k^2} \frac{dk}{2 \pi \omega_k \left[1 + \left(\frac{\omega_0}{2 \omega_k}\right)^2 \right]}.
    \label{eq:s_hat}
\end{equation}

Equations (\ref{eq:reynolds-int}) and (\ref{eq:s_hat}) enable us to compute the contribution of the Reynolds stresses to the stochastic excitation for a given acoustic mode. One can choose to model the turbulent kinetic energy spectrum with a given slope $\alpha$. The choice for this parameter depends on the turbulent regime and is influenced by rotation (see Sect. \ref{sub:turbulent_spectra} for more details). The important physical quantities are the values of the convective rms velocity and wavenumber, usually computed in 1-D stellar modelling codes using the Mixing-Length Theory. 

However, as shown in Figure \ref{fig:interdependencies}, rotation affects convection, which in turn influences the stochastic excitation of modes through the Reynolds stresses source term we compute. Rotation indeed influences convection, leading to a lower (resp. higher) value for $v_c$ (resp. $k_c$) in the Rotating Mixing-Length Theory compared to the non-rotating case. To model the impact of rotation on the stochastic excitation of modes, we thus require a robust model and prescription for rotating convection.

\section{Rotating convection}
\label{sec:rotating_convection}
\subsection{The state of the art}
\par To model rotating convection in stellar or planetary interiors, the most realistic approach is to use direct numerical simulations (DNSs), solving the underlying equations of fluid dynamics. Progress has been made, following the increase of computing power. For instance, the community has computed large sets of Large Eddy Simulations of the 3D non-linear dynamics of stellar rotating convective zones in spherical geometry \citep[we refer the reader to][]{brun_magnetism_2017, smolarkiewicz_eulag_2013, brun_powering_2022, kapyla_simulations_2023}. To tackle specific phenomena, local DNS is also often used instead of global simulations, neglecting the body's curvature and using Cartesian boxes, while pushing the Reynolds number to its maximum possible value \citep[e.g.][]{julien_strongly_1998, julien_statistical_2012, currie_convection_2020, vasil_rotation_2021}. However, these simulations are very demanding and they cannot systematically explore large regions of the parameters space yet. 
\par To study stellar structure and long-term evolution in astrophysics, heat transport by convection is most of the time quantified using MLT \citep[e.g.][]{bohm-vitense_uber_1958, gough_mixing-length_1977}. This unidimensional theory is broadly implemented in stellar structure and evolution codes, such as MESA \citep{paxton_modules_2011, paxton_modules_2013, paxton_modules_2015, paxton_modules_2018, paxton_modules_2019, jermyn_modules_2023}. Despite its simplicity, it has proven to be relevant in modelling stars and planets along their evolution. MLT is based on the maximisation of the heat flux transported by convection, following the principle of \cite{malkus_heat_1954}. The convective MLT length scale and the convective velocity are governed by the linear mode that maximises the convective heat flux. As such, the MLT is thus a local and monomodal theory.

\par However, it omits many important effects, including rotation and magnetic fields. To address these simplifications, a theory of the MLT that takes into account rotation and magnetism has been proposed by \cite{stevenson_turbulent_1979} and extended in the case of rotation by \cite{barker_theory_2014} and \cite{augustson_model_2019}. This R-MLT has not been widely implemented in 1D stellar evolution codes yet, one exception being \cite{ireland_radius_2018} (see also \cite{michielsen_probing_2019} and \cite{dumont_lithium_2021} for the specific study of convective penetration with good agreement with asteroseismology of early-type stars and observations of light-elements abundances in low-mass stars, respectively). \cite{ireland_radius_2018} show that the entropy gradient in low-mass stars can be modified by rotation, leading to structural and evolutionary changes, although such changes are found negligible. The same conclusion has been obtained in the case of giant gaseous planets by \cite{fuentes_rotation_2023} and \cite{fuentes_evolution_2024}. The authors suggest that planetary rotation is another factor contributing to the longevity of primordial composition gradients in Jupiter, as observed by the Juno spacecraft. Furthermore, the prescriptions of \cite{stevenson_turbulent_1979} have been successfully confronted with nonlinear numerical simulations in Cartesian geometry \citep{barker_theory_2014, currie_convection_2020}. In these simulations, the R-MLT formalism established in \cite{stevenson_turbulent_1979} holds well for three decades in Rossby number (defined in Eq. \ref{eq:rossby_def}), which compares the inertia of convective flows to their Coriolis acceleration. The prescription for the convective overshoot with rotation derived by \cite{augustson_model_2019} in the R-MLT framework has also been successfully compared with numerical simulations \citep{korre_dynamics_2021}. Finally, the R-MLT theory has been used to model the stochastic excitation of gravito-inertial stellar oscillation modes by turbulent convection in rapidly rotating stars with predictions in agreement with asteroseismic observations of rapidly rotating early-type stars \citep{neiner_seismic_2012, neiner_astronomy_2020}. This last study shows a high interest in using this theory to study the impact of rotation on the stochastic excitation of stellar oscillation modes in rotating solar-type pulsators.

\subsection{Prescriptions for rotating mixing-length theory}

\par In the present work, we make use of the prescription of \cite{augustson_model_2019} to account for the modification of convection by rotation, generalising the model by \cite{stevenson_turbulent_1979}. This latter provides prescriptions for the asymptotic regimes of slow and rapid rotations while \cite{augustson_model_2019} prescriptions can be applied for any rotation rate (and Rossby number as defined in Eq. \ref{eq:rossby_def}). Both follow the heat flux maximisation principle proposed by \cite{malkus_heat_1954}: the convective flow is dominated and modelled by the mode that transports the most heat.  
\par \cite{stevenson_turbulent_1979} gives a modulation factor in each case. The convective velocity (resp. convective wavenumber) with rotation $u_c$ (resp. $k_c$) is expressed with respect to the convective velocity (resp. convective wavenumber) without rotation $u_0$ (resp. $k_0$): 
 \begin{equation}
\begin{aligned}
u_c = \tilde{U}(\mathcal{R}o) u_0, \\
k_c = \tilde{K}(\mathcal{R}o) k_0,
\end{aligned}
\end{equation}
where we use the effective convective Rossby number introduced by \cite{stevenson_turbulent_1979} computed as a function of the non-rotating characteristic velocity and the convective scale:

\begin{equation}
    \mathcal{R}o = \frac{u_0 k_0}{2 \Omega \cos{\theta}}.
    \label{eq:rossby_def}
\end{equation}
Here $\theta$ is the colatitude. The convective velocity without rotation $u_0$, as well as the convective wavenumber without rotation $k_0$ are outputs from the MESA stellar structure and evolution code (see Sec. \ref{sec:numerical_estimates} for the detailed method and results). 

\begin{figure}[!hb]
    \centering
    \includegraphics[width = \linewidth]{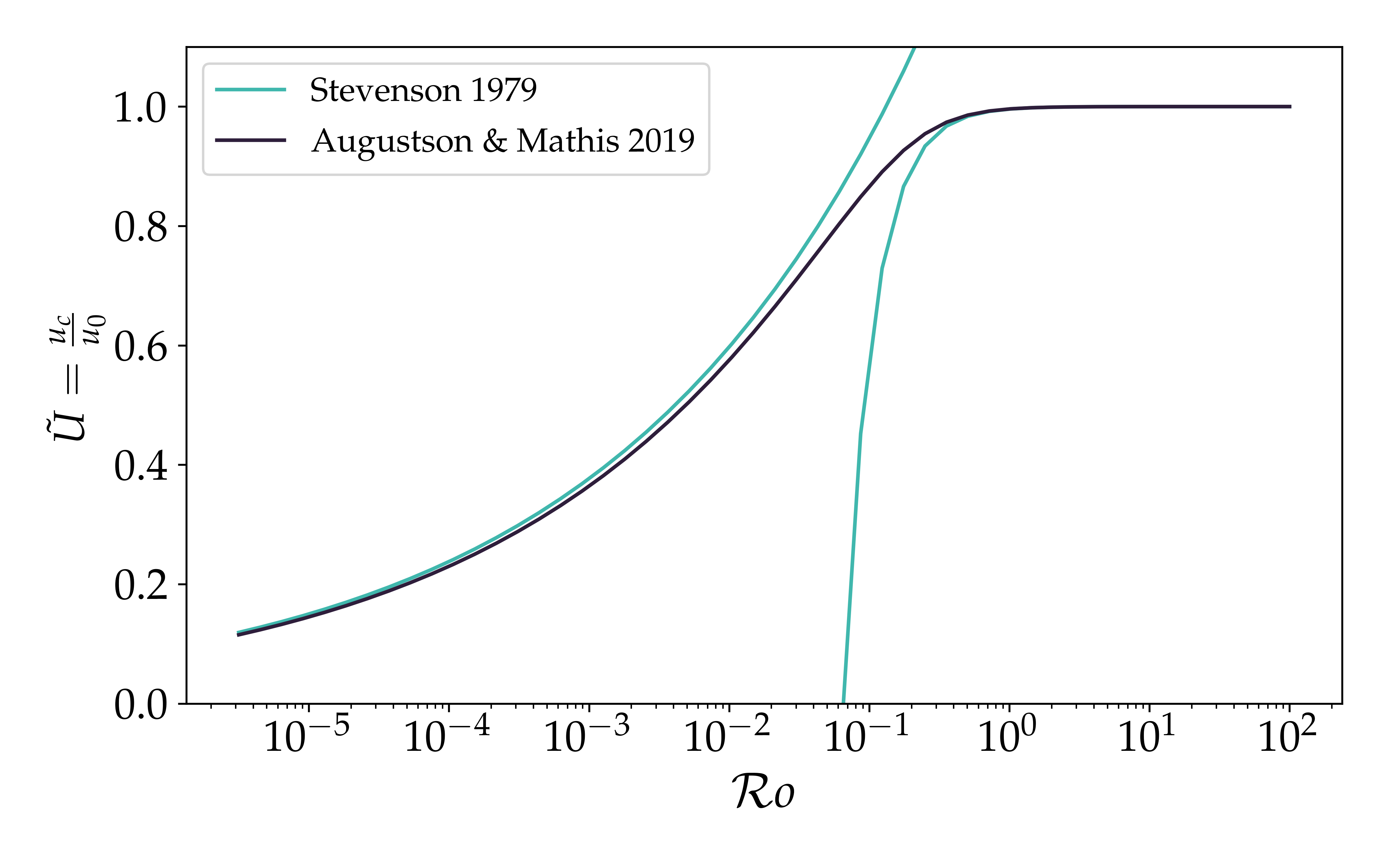}
        \includegraphics[width = \linewidth]{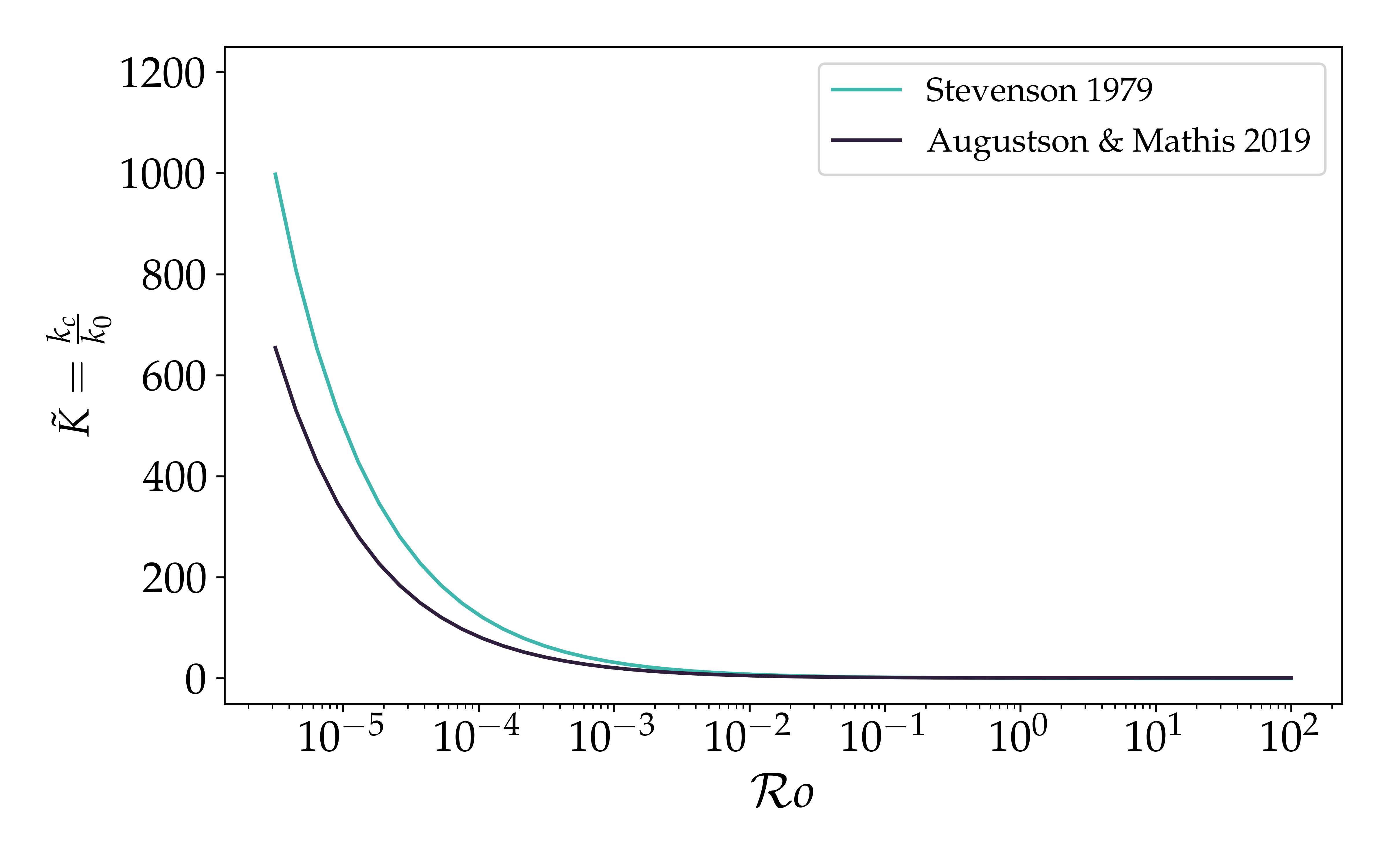}
    \caption{Rotating mixing-length theory: modulation of the convective velocity (\textit{top panel}) and wavenumber (\textit{bottom panel}) in \cite{stevenson_turbulent_1979} (green lines) and in \cite{augustson_model_2019} (dark lines).}
    \label{fig:conv_rot_stev_aug}
\end{figure}

The values of the modulation factors $\tilde{U}(\mathcal{R}o)$ and $\tilde{K}(\mathcal{R}o)$ derived by \cite{stevenson_turbulent_1979} in the slow and rapid rotation limits are given in Table \ref{tab:stevenson_rot}. 

\begin{table}[h!]
    \centering
\renewcommand{\arraystretch}{2}
    \caption{Modulation of the convective velocity and scale by rotation in the rapidly and slowly rotating regimes, respectively, following \cite{stevenson_turbulent_1979}.}
    \begin{tabular}{p{1.5cm}  p{2.5cm}  p{2.5cm}}
    \hline
     \hline
     & $\mathcal{R}o \ll 1$ &  $\mathcal{R}o \gg 1$ \\
     \hline
        $\tilde{U}(\mathcal{R}o)$ & $1.5 \mathcal{R}o^{1/5}$   & $1 - \frac{1}{242 \mathcal{R}o^2}$\\
        
         $\tilde{K}(\mathcal{R}o)$ & $0.5 \mathcal{R}o^{-3/5}$   & $1+ \frac{1}{82 \mathcal{R}o^2}$\\
    \hline  
    \end{tabular}
    \label{tab:stevenson_rot}
\end{table}

\par \cite{augustson_model_2019} uses the same principle, and solves the following equation: 
\begin{equation}
    2 z^5-5 z^2-\frac{18 \cos ^2 \theta}{25 \pi^2 \operatorname{Ro}_{\mathrm{c}}^2 \tilde{s}^2}=0, 
    \label{eq:z_augustson}
\end{equation}
where $\tilde{s}$ is defined as   $ \tilde{s} = 2^{1/3} 3^{1/2} 5^{-5/6} $. 

One can in turn find the velocity modulation and the wavenumber modulation, as in \cite{stevenson_turbulent_1979}: 
\begin{equation}
    \tilde{U} = \frac{5 \tilde{s}}{\sqrt{6}} z^{-1/2},
    \label{eq:u_augustson}
\end{equation}
\begin{equation}
    \tilde{K} = \sqrt{\frac{2}{5}} z^{3/2}.
    \label{eq:k_augustson}
\end{equation}

We refer the reader to the Appendix \ref{sec:appendix_mlt}, where we detailed the method from \cite{augustson_model_2019} for the derivation of Equations (\ref{eq:z_augustson}), (\ref{eq:k_augustson}), and (\ref{eq:u_augustson}). 

\par We compared the output of $\tilde{U}$ and $\tilde{K}$ with these two methods in Fig. \ref{fig:conv_rot_stev_aug}. For the velocity modulation, \cite{augustson_model_2019} predictions are in excellent agreement with those of \cite{stevenson_turbulent_1979} in the asymptotic slowly and rapidly rotating regimes. When it comes to the wavenumber modulation for the low Rossby number case, the exponent is in agreement, only the prefactor differs between \cite{stevenson_turbulent_1979} and \cite{augustson_model_2019}. As both authors consider a linear dispersion relation for the convective instability, there can subsist a difference in the prefactor. The Rossby number in Solar-like stars is superior to $\mathcal{R}o \geq 10^{-3}$ (see e.g. Fig. \ref{fig:spheres_soleil}): in this regime the difference between \cite{stevenson_turbulent_1979} and \cite{augustson_model_2019} formulations is negligible. In the following, we estimate the power injected into stellar acoustic modes by turbulent convection while taking its modification by rotation into account making use of the most recent and non-asymptotic \cite{augustson_model_2019} prescriptions.

\subsection{Turbulent spectrum in rotating turbulence}
\label{sub:turbulent_spectra}

\par Rotation also affects the turbulent kinetic energy spectrum defined in Eq. (\ref{eq:energy_normalised}), which in turn influences the stochastic excitation of acoustic modes (see Fig. \ref{fig:interdependencies}). Numerical simulations and experiments have highlighted a wide panel of behaviours in rotating convection or free turbulence. Theoretical, experimental, and numerical studies provide scalings for the kinetic energy spectrum \citep{julien_heat_2012}. \cite{zhou_phenomenological_1995, mahalov_analytical_1996} propose $E(k) \propto k^{-2}$, in a phenomenological study. When the turbulence is forced at a scale $k_f$, \cite{smith_transfer_1999} find with numerical simulations $E(k) \propto k^{-3}$ for $k<k_f$, and $E(k) \propto k^{-5/3}$ for $k> k_f$. \cite{mininni_rotating_2010} also find a scaling $E(k) \propto k^{-5/3}$, similar to the Kolmogorov spectrum scaling for forced turbulence. Nevertheless, rotating turbulence is strongly anisotropic: for a more accurate description of the turbulent cascade, one must generally distinguish the wavenumber parallel to the rotation axis $k_{\parallel}$, and the orthogonal component $k_{\perp}$. We neglect this anisotropy in the present study as a first step because we are considering the isotropic formalism of R-MLT. Following Eq. (\ref{eq:energy_normalised}), the turbulent kinetic energy spectrum scales like $E(k) \propto k^{-\alpha}$ and we explore the dependency of the stochastic excitation with the slope $\alpha$ of this spectrum. More details regarding the turbulent spectra in rotating turbulence can be found for example in \cite{godeferd_structure_2015, alexakis_cascades_2018}.

\begin{figure}[!hb]
    \centering
\includegraphics[width=0.95\linewidth]{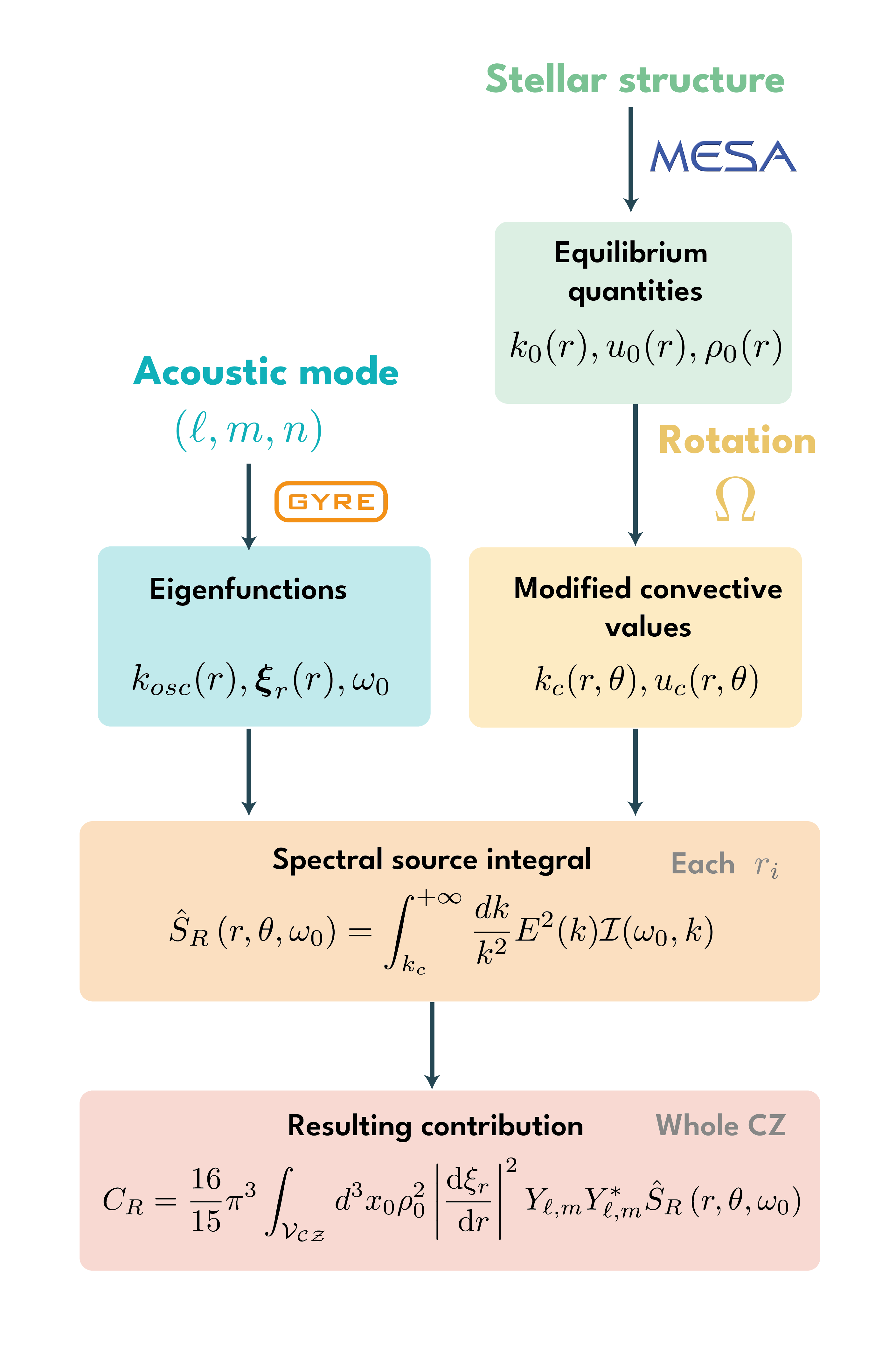}
    \caption{Method used to assess the power injected into stellar acoustic modes. We first computed a stellar model with MESA, where the modification of the stellar structure by rotation is not taken into account. For a given value of rotation $\Omega$, we compute the local Rossby number and then the modified convective wavenumber and velocity, following the prescriptions by \cite{augustson_model_2019}. We compute the eigenfunctions and eigenfrequencies for the acoustic modes using GYRE. It enables us to compute at each radius position the local spectral source term $\hat{\mathcal{S}}(r, \theta)$. Finally, we integrate the resulting contribution on the whole Convective Zone (CZ).}
    \label{fig:method_compute}
\end{figure}

\section{Results: Impact of rotation on the excitation of acoustic modes in solar-like pulsators}
\label{sec:numerical_estimates}
\begin{figure*}[!ht]
    \centering
    \includegraphics[width=0.99\linewidth]{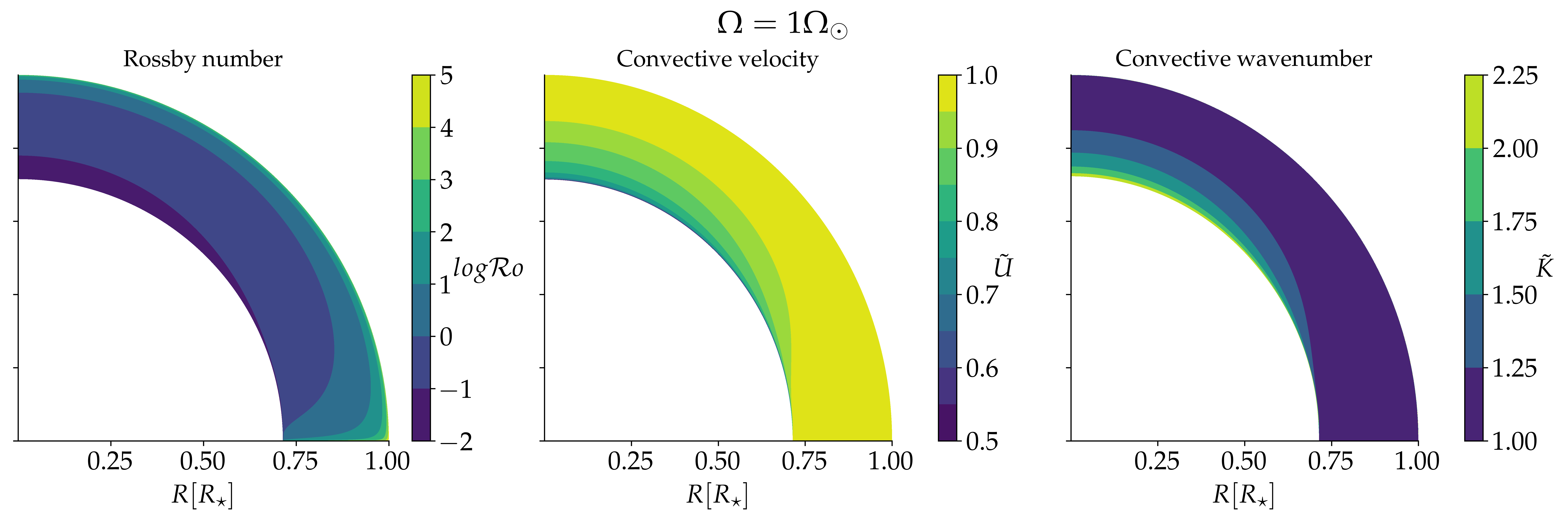}
    \caption{Solar model computed with MESA. (Left) Effective Rossby number as defined by \cite{stevenson_turbulent_1979} in Eq. (\ref{eq:rossby_def}) in the Sun's convective zone. The region at the base of the convective zone and at the poles have the lowest Rossby number, then is the most affected by rotation. (Centre) The same figure, but for the modulation of the convective velocity in the framework of R-MLT of \cite{augustson_model_2019}. The convective velocity is diminished by $50\%$ in the lowest Rossby number region, at the base of the convective zone. (Right) Modification of the convective wave number with rotation, following \cite{augustson_model_2019}. In the lowest Rossby number zone, the size of the convective eddies is the most diminished.}
    \label{fig:spheres_soleil}
\end{figure*}

In this section, we assess the power injected by the rotating turbulent convection into the acoustic modes, using the theoretical formalism expounded upon above. We use a Solar-like star 1D model, computed with the stellar evolution code MESA for quantities like sound speed, convective velocity or density profiles. We use in combination the stellar oscillation code GYRE to compute stellar acoustic modes eigenfunctions $\bm{\xi}$ and their eigenfrequencies. 
For each mode, we use the following methodology: 
\begin{enumerate}
    \item For a given rotation rate $\Omega$, we compute the effective Rossby number using a profile from MESA. Next, we calculate the convective velocity modified by rotation, following the prescription from \cite{augustson_model_2019}. We refer the reader to Appendix \ref{sec:appendix_mlt} for more details. Fig. \ref{fig:spheres_soleil} shows the Rossby number, modified convective velocity, and modified wavenumber for a Sun-like model. Since the effective Rossby number diverges near the equator, we only display points where $\theta < \pi/2$.
    \item We compute the oscillations eigenfunctions $\xi_r(r)$ and eigenfrequencies $\omega_0$ with GYRE.
    \item At each radial position, we compute the integral $I(r,\omega_0)$ from Eq. (\ref{eq:i_omega}) over all the eddies scales. 
    \item Finally, we integrate over the whole star's convective envelope, using Eq. (\ref{eq:reynolds-int}).
\end{enumerate}
This process, which is summarised in Fig. \ref{fig:method_compute}, is then repeated for any stellar model, for different modes $(\ell, m, n)$ and values for the rotation $\Omega$. 
In the following work, we take the average surface rotation of the Sun as a reference: $\Omega_{\odot} = 28 \textsc{  } \rm days$ \citep{thompson_internal_2003, garcia_tracking_2007}.

\subsection{Influence of the turbulent eddy-time correlation}
As explained in Section 2, one can choose either a Lorentzian or a Gaussian turbulent time-correlation function. First, we examine the impact of either of these choices on mode excitation. Using \cite{augustson_model_2019} model for rotating convection, the lower integration bound $k_c$ in $\hat{\mathcal{S}}(r, \theta)$ increases when the rotation increases. However, only a Lorentzian time-correlation function is significantly modified by rotation, the Gaussian time-correlation function being too steep for low-$k$ values. We plotted in Figure \ref{fig:lorentz-gauss} $\mathcal{J}(k)$ the integrand of $\hat{\mathcal{S}}(r, \theta)$,
\begin{equation}
\hat{S}_R\left(r, \theta, \omega_0\right)=\int_{k_c}^{+\infty} \underbrace{\frac{E^2(k)}{k^2}  \mathcal{I}(\omega_0,k)}_{\mathcal{J}(k)} dk,
\label{eq:s_hat_integrand}
\end{equation}
\noindent to underline this effect. As only a Lorentzian eddy-time correlation function significantly impacts the stochastic excitation of acoustic modes in the presence of rotation, we adopt this choice in the following semi-analytical study. Indeed, \cite{mathur_revisiting_2019} have shown that mode detections are related to stellar rotation periods. Therefore, we assume that rotation influences the stochastic excitation of acoustic modes, supporting the choice for a Lorentzian eddy-time correlation function in this framework.

Choosing a Lorentzian eddy-time correlation function, we examine the impact of rotation on the various acoustic modes. We use a Sun-like model, with metallicity $Z=0.02$ (see Appendix \ref{sec:inlist} for the MESA inlists). Here, we consider a fixed value for the turbulent spectrum slope: $\alpha = -5/3$ \citep{kolmogorov_dissipation_1941}. For each mode, we compute the power $\mathcal{P}_0$ injected into the oscillations without rotation ($\Omega=0$), and the injected power $\mathcal{P}_{\Omega}$ taking rotation into account. 
\par
\begin{figure*}[h]
    \centering
\includegraphics[width=0.49\linewidth]{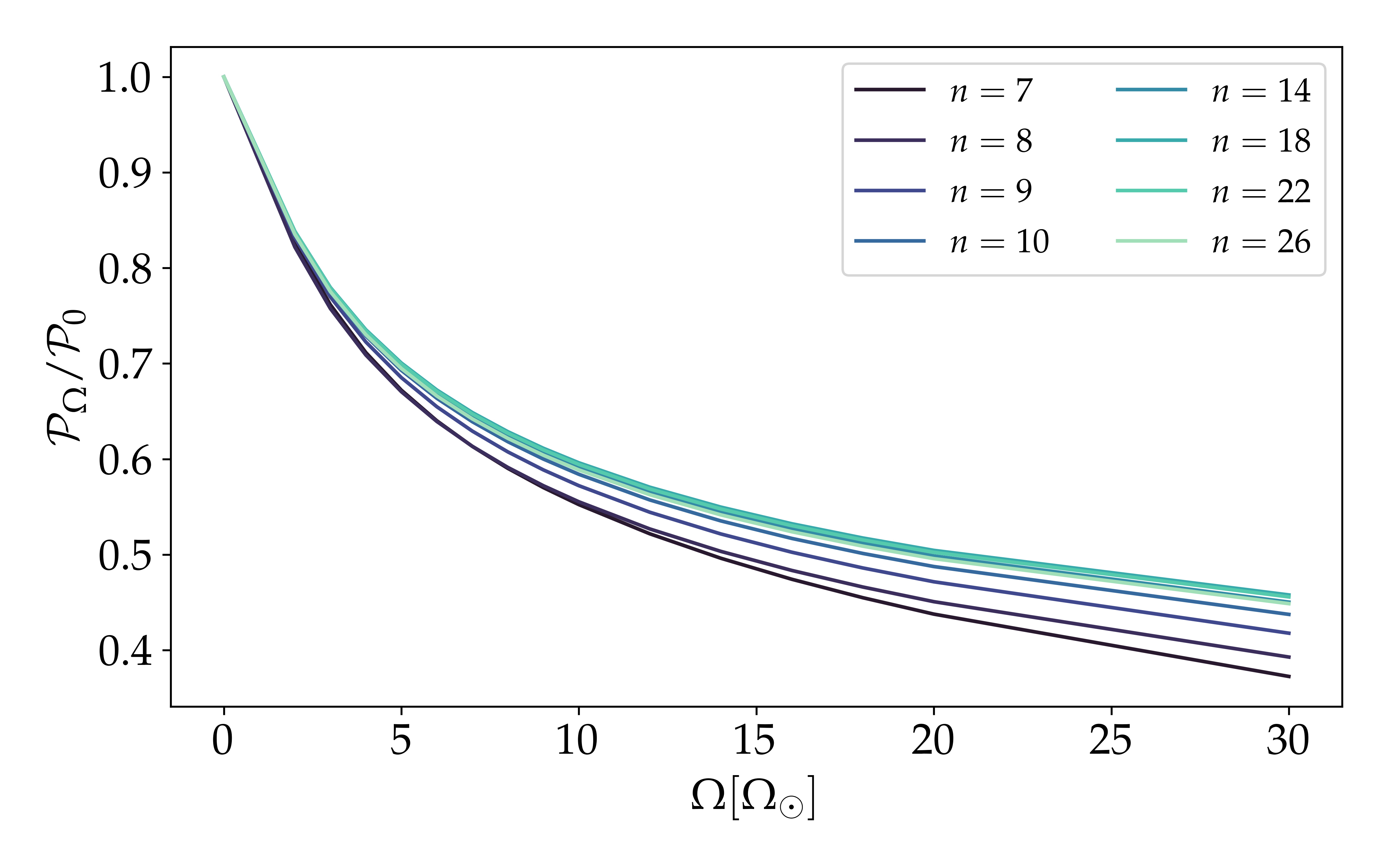}
\includegraphics[width=0.49\linewidth]{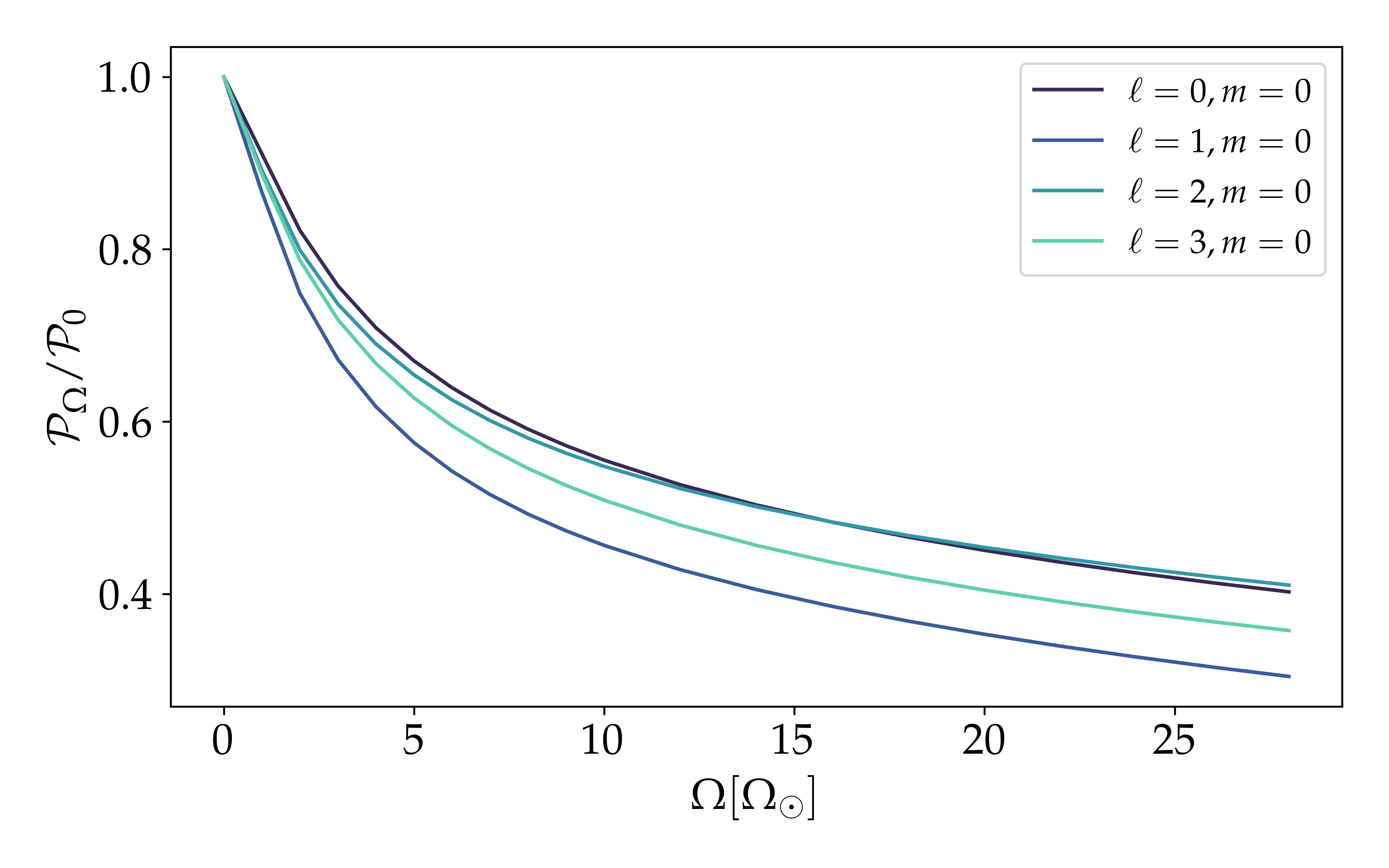}
\caption{Impact of rotation on different acoustic modes. (Left) Influence of the radial order $n$ for the modes with $\ell=0, m=0$: the modes with the lowest values of $n$ are most influenced by rotation. (Right) Influence of the horizontal degree $\ell$ for the modes with $n=8$ and $m=0$.}
\label{fig:puissance_modes_rotation}
\end{figure*}

 Figure \ref{fig:puissance_modes_rotation} shows the ratio $\mathcal{P}_{\Omega}/\mathcal{P}_0$ as a function of rotation for different modes. As expected, rotation tends to inhibit the excitation: the higher the rotation rate, the lower the power injected by the stochastic excitation. Furthermore, rotation does not affect the modes equally: for a fixed value of $\ell=0$, the lower the $n$ order, the more inhibited the modes (see the left panel in Fig. \ref{fig:puissance_modes_rotation}). In Eq. (\ref{eq:reynolds-int}), the term $\hat{\mathcal{S}}$ diminishes when the frequency $\omega_0$ increases. However, when computing the modes using GYRE the remaining term with $d\xi_r/dr$ makes the overall injected power increase. This dependence of the injected power on the order and degree is mostly due to the term in $\xi_r$ in Eq. (\ref{eq:reynolds-int}), which is computed with GYRE here. Indeed modes with higher eigenfunction amplitudes at the base of the convective zone, where the Rossby number is smaller, are more strongly influenced by rotation, leading to a more pronounced reduction in the injected power. 

For modes with increasing $\ell$ and a fixed $n = 8$ and $m = 0$, the impact of rotation on mode excitation does not follow a clear trend. This non-monotonic variation with $\ell$ arises from the role of the spherical harmonics in Eq. (\ref{eq:reynolds-int}), which modulate the excitation differently at each latitude. Modes with $\ell=0$ and $\ell = 2$ are almost equally influenced by rotation, while modes with $\ell =1$ and $\ell = 3$ are more affected (see the right panel in Fig. \ref{fig:puissance_modes_rotation}). 

\begin{figure*}[h]
    \centering
    \includegraphics[width = 0.33 \linewidth]{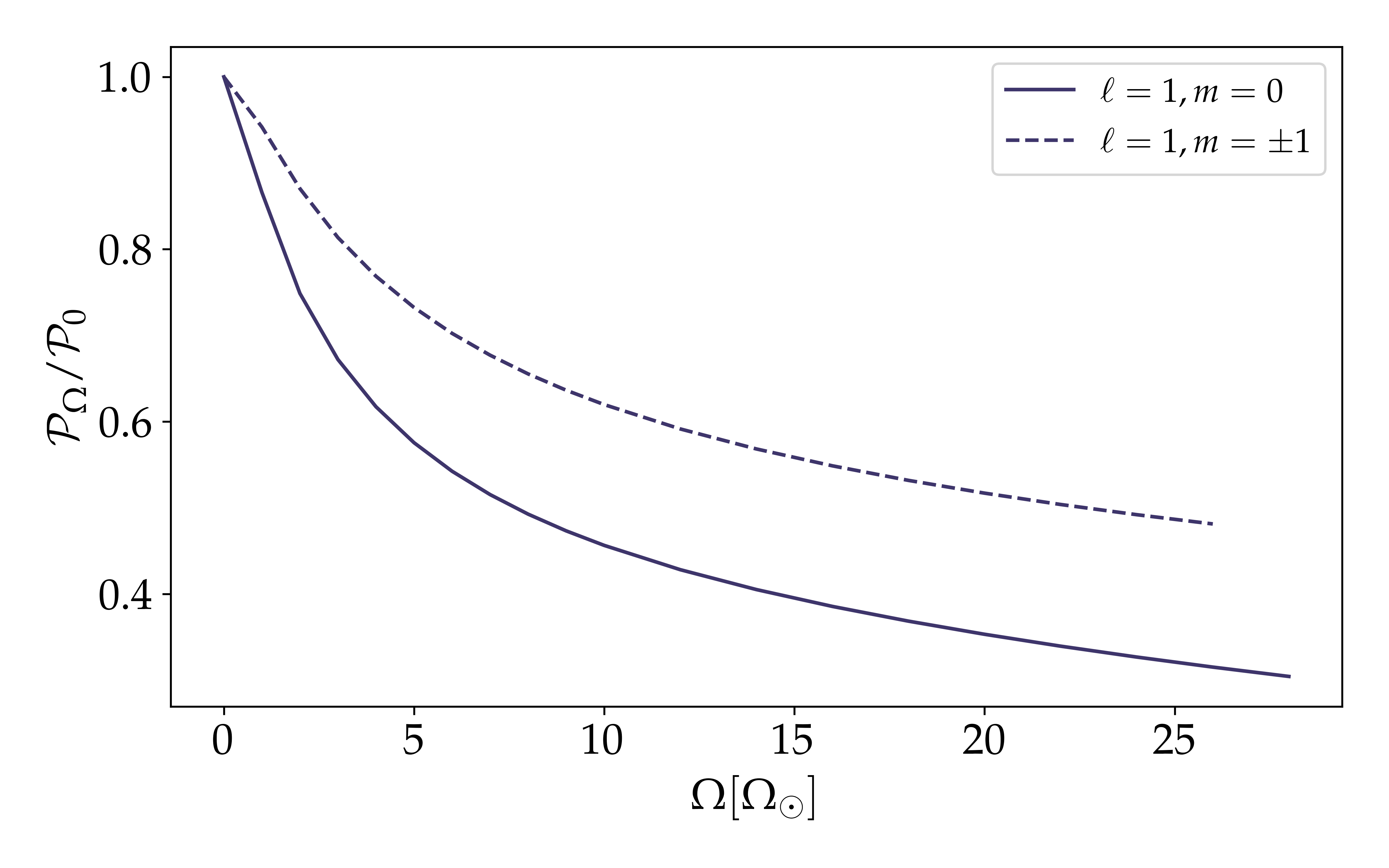}
    \includegraphics[width = 0.33 \linewidth]{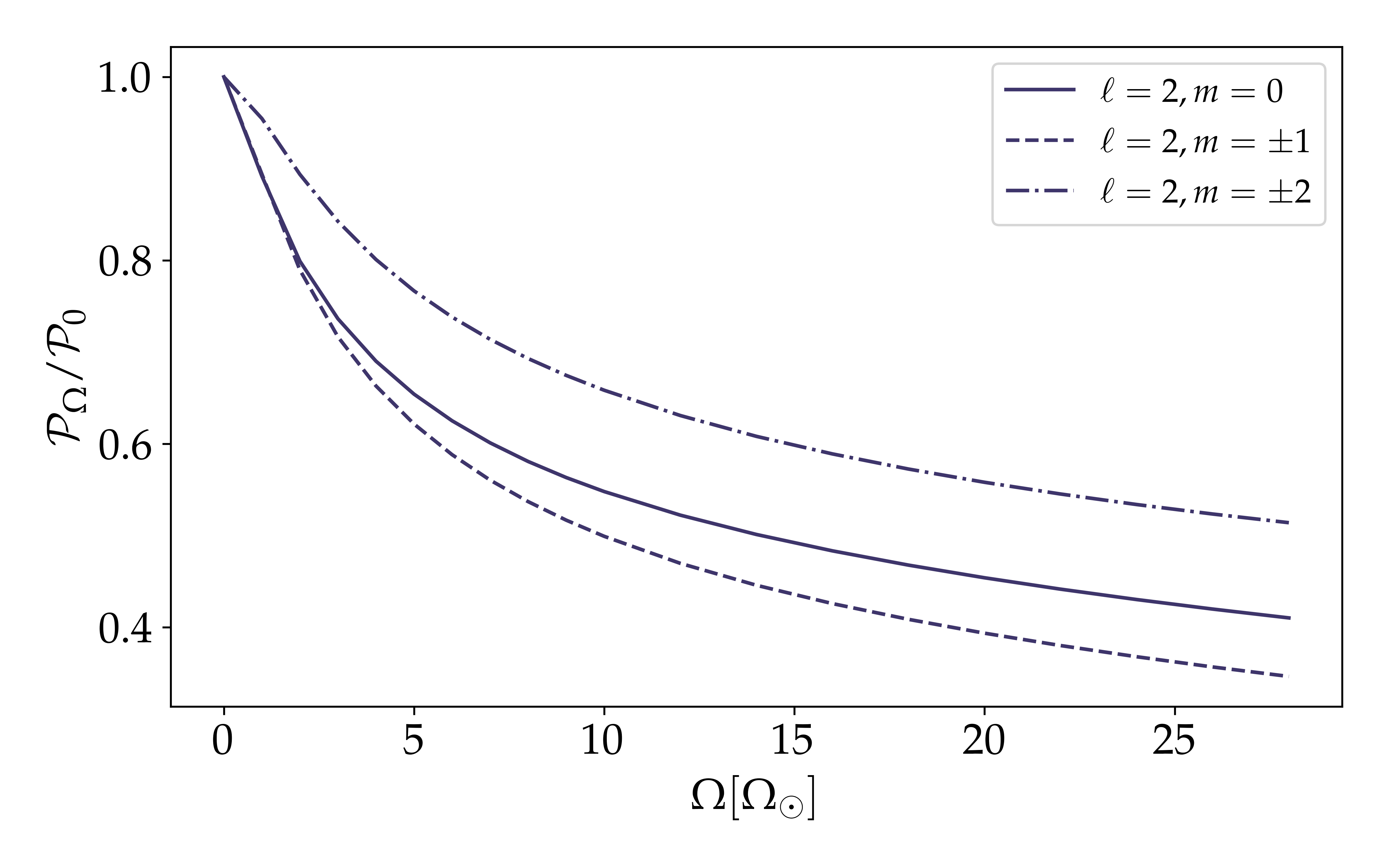}
    \includegraphics[width = 0.33 \linewidth]{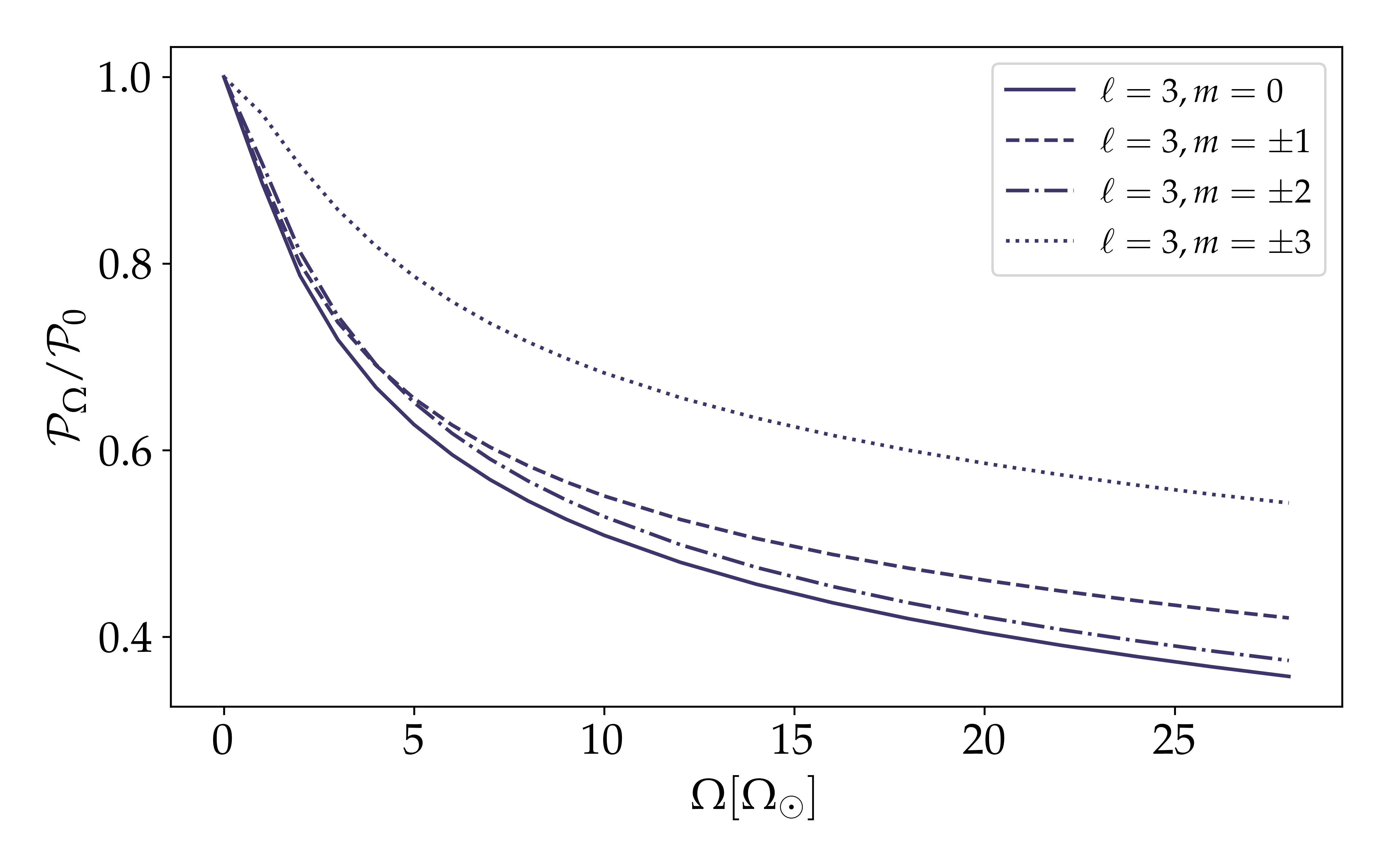}
    \caption{Influence of rotation on the excitation power for different values of azimuthal order $m$ for a Sun-like stellar model. (Left) $\ell=1, n=8$ modes with different $m$. (Center) $\ell=2, n=8$ modes with different $m$ (Right) $\ell=3, n=8$ modes with different $m$.}
    \label{fig:comparaison_m}
\end{figure*}

Moreover, the excitation also depends on the azimuthal number  $m$, which appears in spherical harmonics within the Reynolds stresses contribution in Eq. (\ref{eq:reynolds-int}). As the spherical harmonics modulus verify $Y_{\ell,m}^{*} = (-1)^m Y_{\ell,-m}$, the same values for $\lvert m \rvert$ give the same power injected in the oscillations in this framework. We studied the influence of $m$ and $\ell$ for a fixed $n=8$ radial order in Fig. \ref{fig:comparaison_m}. There is no direct tendency observed: $(\ell = 2, m= \pm 1)$ mode is more diminished by rotation than the corresponding $m=0$ mode, whereas for $\ell = 1$ and $\ell=0$, values of $m \neq 0$ are always less diminished by rotation than the corresponding $m=0$ modes.

\subsection{Influence of the turbulent spectrum}
\par
\begin{figure}[h]
    \centering
\includegraphics[width=\linewidth]{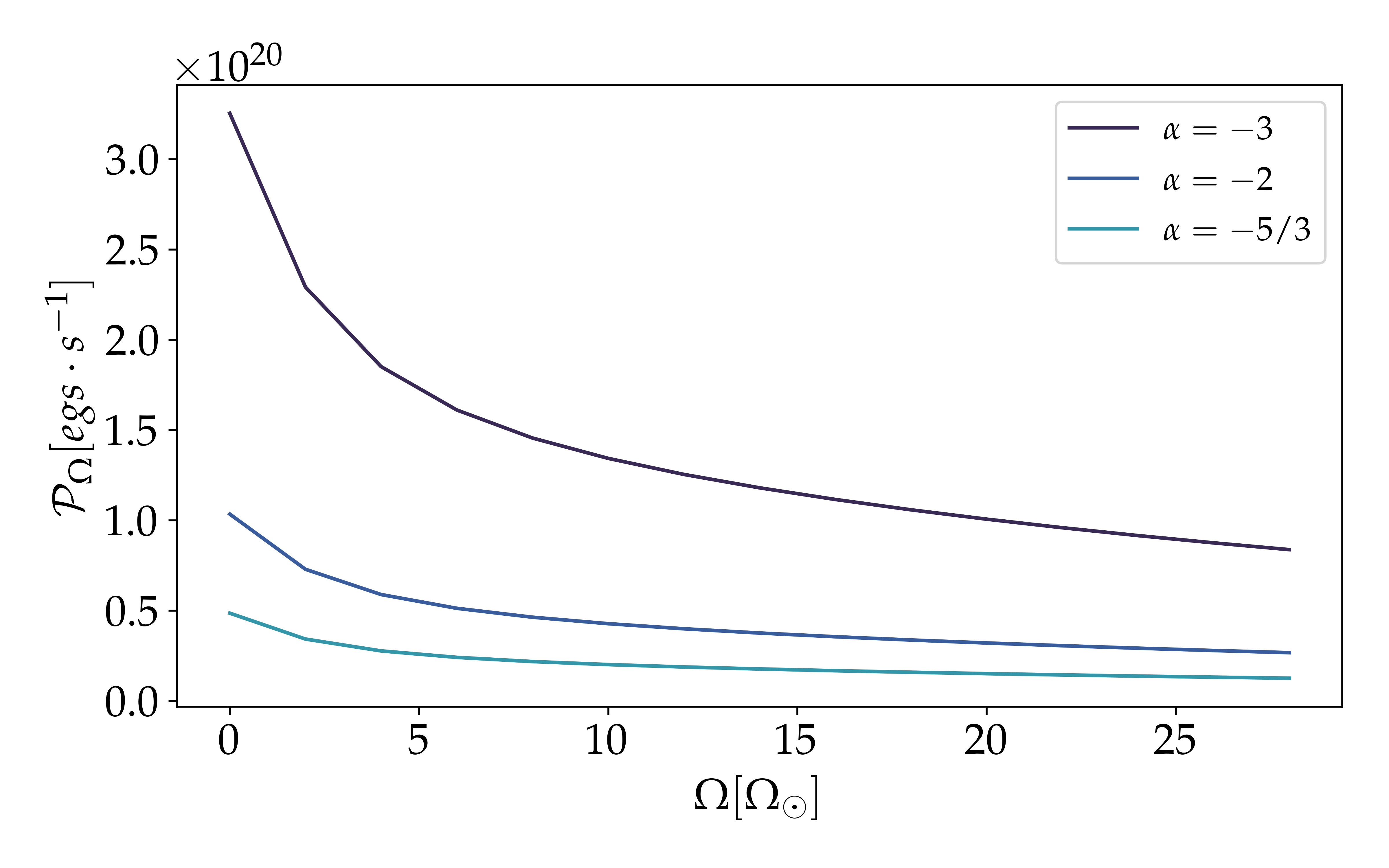}
    \caption{Influence of the choice for the kinetic energy spectrum in the present theoretical model. The steeper the slope $\alpha$ for the kinetic energy spectrum, the higher the power injected into the modes.}
    \label{fig:turbulence}
\end{figure}

 In the previous section, we chose a value of $-5/3$ \citep{kolmogorov_dissipation_1941} for the turbulent spectrum, in Eq. (\ref{eq:s_hat}). However, rotation also affects the turbulent kinetic energy spectrum (see Fig. \ref{fig:interdependencies} and Section \ref{sub:turbulent_spectra}). We compute the injected power into the modes for different values of the turbulent spectrum in Fig. \ref{fig:turbulence}. %

We chose the values found in the literature: $\alpha = -5/3$ \citep{kolmogorov_dissipation_1941, mininni_rotating_2010}, $\alpha = -2$ \citep{zhou_remarks_1998}, and $\alpha = -3$ \citep{smith_transfer_1999}. 
We used the same method as before, focusing on the mode $\ell=0$, $n=8$, which is strongly influenced by rotation as explained in the previous section. The power injected into the modes is influenced by the choice for the turbulent spectrum scaling: the steeper the slope, the higher the injected power. Indeed, we can show from Eq. (\ref{eq:s_hat}) that
\begin{equation}
    \hat{S}_R(r, \theta, \omega_0) \sim \frac{(\alpha -1)^2}{\sqrt{1 - 2^{1-\alpha}}} \int \frac{dK}{K^{(7 + 3 \alpha)/2} + A (1 - 2^{1-\alpha}) K^{(5 \alpha+1)/2}},
\end{equation} where $A$ is a prefactor that does not depend on $\alpha$. Numerically, we find that the higher the value of $\vert \alpha \rvert$, the higher the integrand, which leads to more power injected by the stochastic excitation 

\subsection{Impact of the stellar structure}
All the studies presented in the previous parts were carried out using the same solar-like star model computed with MESA. We now focus on other stars with a different mass or metallicity to explore the effects of stellar structure on our results. We thus apply the same algorithm to five different stellar models. 

First, we study the impact of the stars' metallicity on the stochastic excitation. We consider Sun-like stars with $1 M_{\odot}$ mass, $4,60 \rm Gyr$ age, and different metallicities $Z=0.002$, $Z=0.02$ and $Z=0.04$, respectively. 

Second, we model a Pre-Main Sequence (PMS) star with a $1 \mathrm{M}_{\odot}$ mass, $50 \rm Myrs$ age and a $Z=0.02$ metallicity.  Indeed, PMS stars are of growing interest in asteroseismology, as they could help to investigate the formation and early evolution phases of stars \citep[see e.g.][]{zwintz_pre-main_2022}. Progress has been made to develop novel ways to detect solar-like oscillations in these stars, despite high background levels \citep{mullner_searching_2021}. In addition, these stars tend to rotate faster than Main Sequence stars \citep[e.g.][]{gallet_improved_2015}, which could further hinder acoustic mode detection.

We also model a young Sun F-type $1.15 M_\odot$ star from the \textit{Kepler} sample, KIC 10644253. Its properties are presented in Table \ref{tab:kic}, based on the work of \cite{salabert_magnetic_2016}. This star is of keen interest as it is a typical F-type star, which has a thinner convective layer than G-type stars and a more rapid rotation. This more rapid rotation characterises F-type stars as they are less braked than G-type stars \citep{garcia_study_2014, breton_rooster_2021, santos_surface_2021}.

\begin{table}[h!]
    \caption{Parameters of the star KIC 10644253: Comparison between the observations and the computed MESA model, with the source of the observed values. }
    \label{tab:kic}
    \begin{tabular}{c|c|c|c}
         Parameter & Observations & Source & Model \\
         \hline
         \hline
        Mass ($M_\odot$) & $1.13 \pm 0.05$ &  \cite{metcalfe_properties_2014} &$1.15$ \\
        Radius ($R_\odot$) & $1.108 \pm 0.016$ & \cite{metcalfe_properties_2014} & $1.120$ \\
        Z ($dex$) & $0.026 \pm 0.05$ & \cite{bruntt_accurate_2012} &$0.026$ \\
        $\log g$ (\textit{dex})  & $4.40 \pm 0.3$ & \cite{bruntt_accurate_2012} &$4.39$ \\
        Age ($Gyr$) & $1.07 \pm 0.25$ & \cite{metcalfe_properties_2014} & $1.32$ \\
        $T_{\rm eff}$ ($K$) & $6030 \pm 60$ & \cite{bruntt_accurate_2012} &$6005$ \\
        $\Omega$ ($\Omega_\odot$) & $2.47 \pm 0.05$ &\cite{garcia_impact_2014} &$-$ \\     
    \end{tabular}
\end{table}

The evolutionary paths of those models in the HR diagram are shown in Fig. \ref{fig:hr_diagram}. Their radial profiles of density, convective velocity, and convective wavenumber in the MLT framework are plotted in Fig. \ref{fig:profiles}. These profiles are obtained using stellar models computed with MESA, where rotation does not modify the structure.
\par
\begin{figure*}[ht]
    \centering
\includegraphics[width=0.49\linewidth]{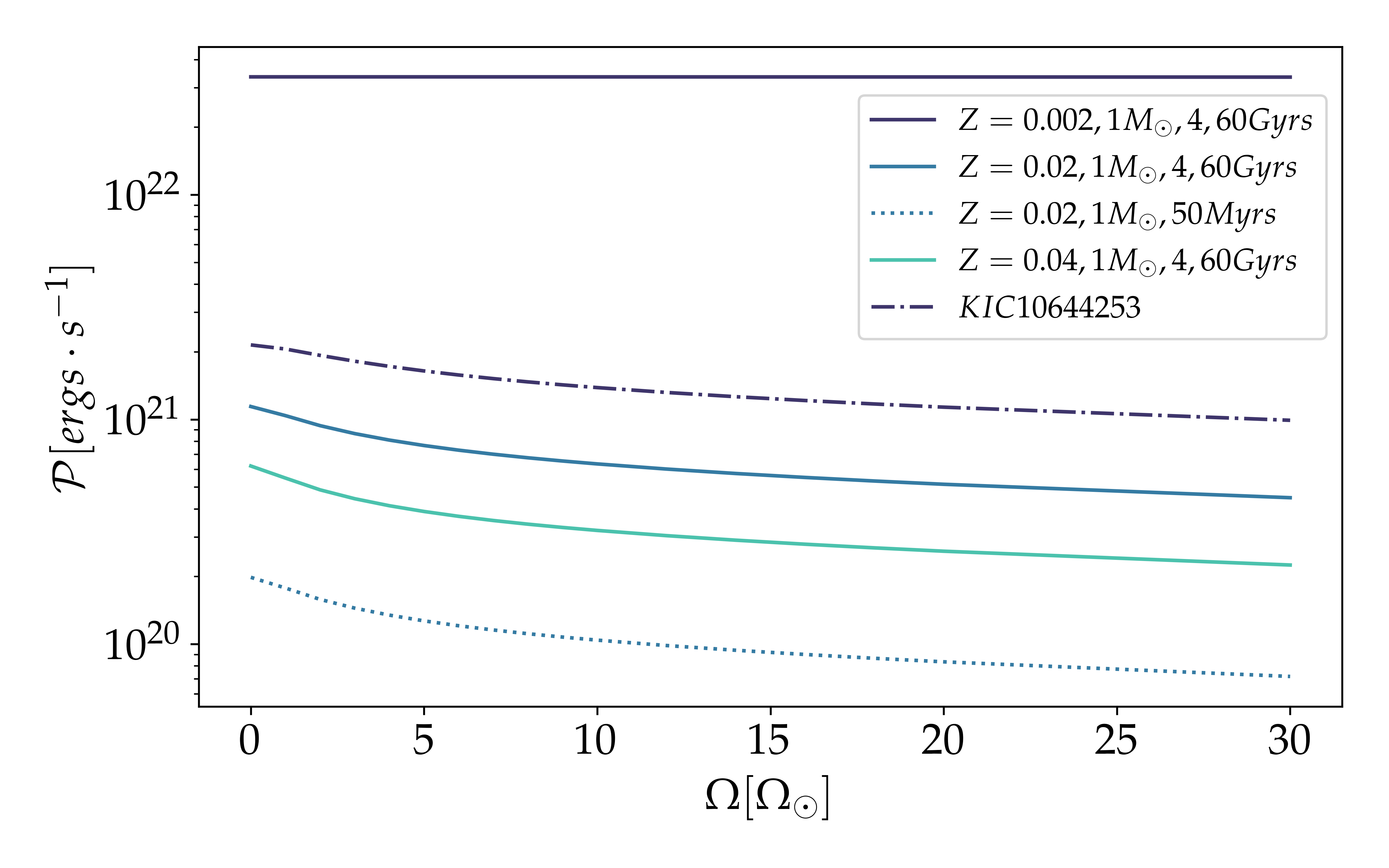}
\includegraphics[width=0.49\linewidth]{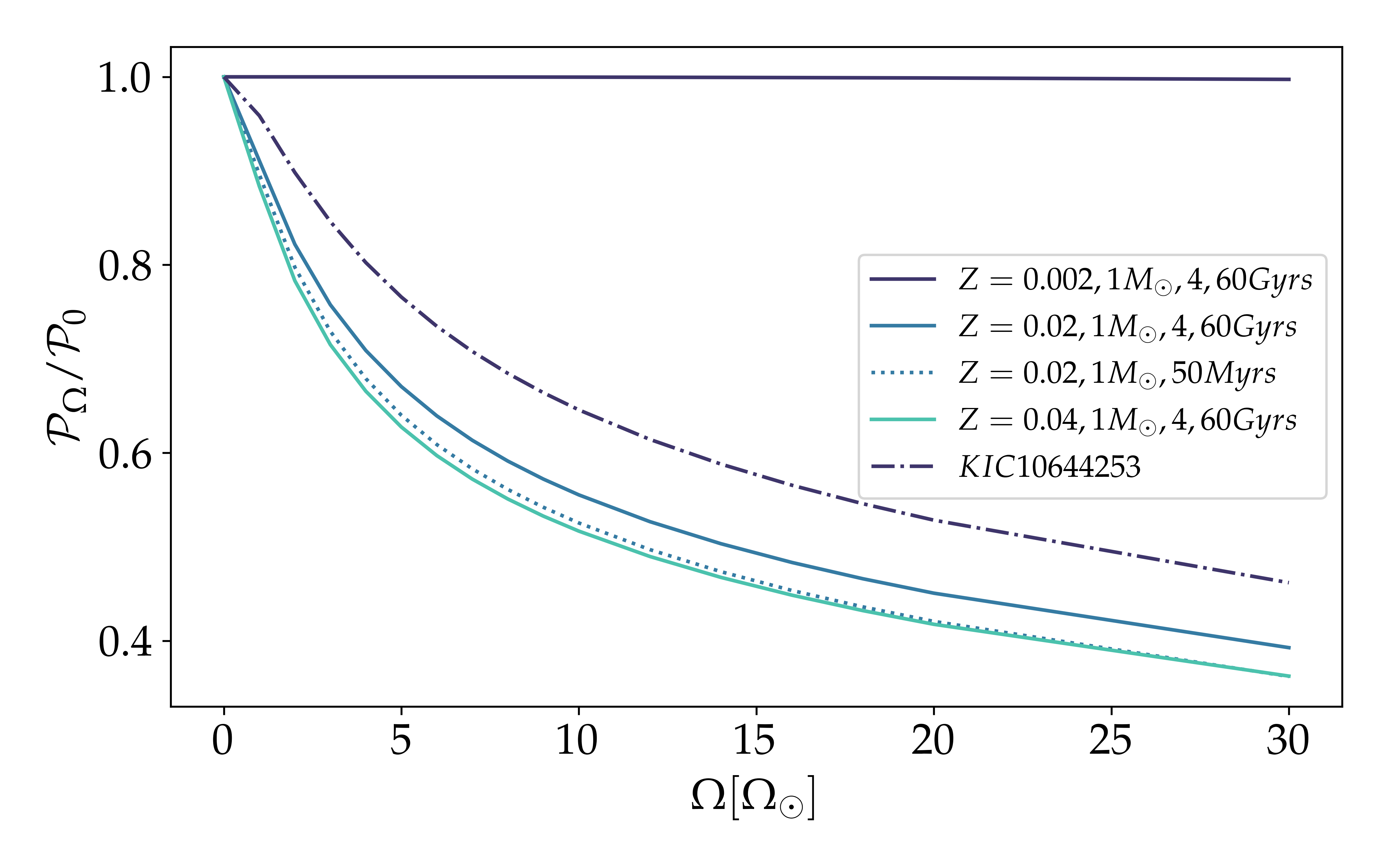}
    \caption{Influence of the stellar structure on the stochastic excitation of acoustic mode $(\ell=0,n=8)$ with rotation. (\textit{Left}) One can notice a stronger decrease in the ratio $\mathcal{P}/\mathcal{P}_0$ for stars with a higher metallicity. Modes in such stars are less affected by rotation. (\textit{Right}) The normalised value of the total power injected $\mathcal{P}$ is plotted.}
    \label{fig:excitation_structure}
\end{figure*}

 Figure \ref{fig:excitation_structure} shows the power injected into the acoustic mode $(\ell=0, n=8)$ for those five different MESA models. First, one notices that the lower the metallicity, the less influence rotation has on the mode amplitudes. For the metal-poor model with $Z=0.002$, the convective zone is thinner than for the other models (see Fig. \ref{fig:profiles}), and the Rossby number is always superior to unity. Thus, the influence of rotation on convection is negligible throughout the whole convective zone, and excitation in this model seems insensitive to rotation. Second, the overall stellar structure plays an important role in the acoustic mode excitation by convection. Indeed, the power injected into the lowest-metallicity model ($Z=0.002$) exceeds the power injected into the $Z=0.02$ model by nearly two orders of magnitude. Although the $Z=0.002$ model has a thin convective zone compared to the $Z=0.02$ model, its convective velocity at the top of the convective zone is higher (see Fig. \ref{fig:profiles}). Most of the power from stochastic excitation is injected at the top of the convective zone, just below the photosphere. The difference in convective velocity at this location thus explains the variation of the injected power. A direct observational link between \textit{Kepler} stars' metallicities and their acoustic mode amplitudes is difficult to emphasise, as other stellar parameters such as luminosity and effective temperature are also influenced by metallicity. \cite{kjeldsen_amplitudes_2011} suggested an empirical scaling relation predicting solar-like oscillation amplitudes. For stars with the same mass and equally damped oscillations, a higher stellar luminosity and a lower effective temperature lead to higher amplitudes. This appears to agree with the present study, although we did not take into account the effects of damping. \cite{mathur_revisiting_2019} searched for a tendency among the metallicities of non-oscillating stars in their \textit{Kepler} sample, using both the Apache Point Observatory Galactic Evolution Experiment survey \citep[APOGEE DR14, ][]{majewski_apache_2017, holtzman_apogee_2018} and the DR2 Large Sky Area MultiObject Fiber Spectroscopic Telescope survey \citep[LAMOST, ][]{de_cat_lamost_2015, luo_vizier_2016}. They noticed a high discrepancy between the values of the two different surveys, leading to uncertain conclusions. One also has to highlight that examining the impact of metallicity has to be handled carefully: not only the metallicity, but the whole stellar structure parameters (such as the effective temperature and the luminosity) play a significant role in mode excitation by convection.  For example, \cite{samadi_corot_2010} found that for a fixed effective temperature metal-poor stars tend to have less excited modes. As in the present work, we compare stellar models with the same mass and age, thus there is no contradiction with the previous study of \cite{samadi_corot_2010}. 

\noindent For the $1 M_{\odot}$ PMS model of metallicity $Z=0.02$ at age $50 \rm Myrs$, nearly ten times less power is injected into the modes than for the corresponding main sequence model (Fig. \ref{fig:excitation_structure}, left panel). For this PMS star, the injected power decreases more rapidly when rotation increases compared with the main sequence model (Fig. \ref{fig:excitation_structure}, right panel). This result is important for selecting relevant targets for PMS pulsating Solar-like stars.

\noindent When it comes to KIC 10644253 ($1.15 M_\odot, Z=0.026$), more power is injected into the modes than into the $Z=0.02$ model, although its metallicity is higher. Indeed, the stellar structure is different in this case with a higher luminosity. Such results corroborate that one has to take into account precisely the whole stellar structure to assess the power injected by the stochastic excitation into the modes for a given star.

\section{Conclusion and perspectives}
\label{sec:conclusion}
\subsection{Discussion}
\par In this work, we generalise the formalism modelling of the stochastic excitation of stellar acoustic oscillation modes by the bulk turbulence in uniformly rotating convective stellar regions. We build on the work by \cite{samadi_excitation_2001} and \cite{belkacem_mode_2009}, taking the modification of convection by rotation into account. To do so, we use the prescription of the Rotating Mixing-Length Theory \citep{stevenson_turbulent_1979, augustson_model_2019}, which is a monomodal approach considering the mode that transports the most heat. 
\par We applied this theoretical formalism to stellar models, using a combination of MESA and GYRE numerical codes. This semi-analytical approach enables us to obtain the following conclusions.
\par First, the choice for the eddy-time-correlation function is paramount. As shown in Sec. \ref{sub:turbulent_spectra}, a Lorentzian eddy-time correlation function significantly impacts the stochastic excitation of acoustic modes in the presence of rotation, while for a Gaussian function, there is no impact on the injected power. Studying the influence of the rotation on the mode amplitudes from an observational point of view can help better constrain the time-correlation spectrum in the convective zone \citep[for more details about the previous works, we refer the reader to the review in][]{samadi_stellar_2015}. Noticing modulations in the mode amplitudes as a function of rotation as reported by \cite{mathur_revisiting_2019} corroborates a Lorentzian function for $\chi_k$ according to our model, in agreement with \cite{samadi_numerical_2003} and \cite{philidet_interaction_2023}. Indeed, \cite{mathur_revisiting_2019} have shown that the mode detection are related to the rotation periods, as illustrated in Fig. \ref{fig:mathur}. A systematic study of mode amplitudes in stars with surface rotation period $P_{\rm rot}$ determination will help confirm it. It constitutes a key constraint on the coupling between turbulent convection and stellar oscillation modes in rotating stars.

\par Second, rotation significantly diminishes the power injected into acoustic modes by turbulent convection. In a realistic solar-like star, the injected power can decrease by up to $60 \%$ for a $20 \Omega_\odot$ rotation rate, compared to the non-rotating case. This is an indirect effect, due to the impact of rotation on convection that we modelled using the Rotating Mixing-Length Theory \citep{stevenson_turbulent_1979}. This is in agreement with the \textit{Kepler} observations of rapidly rotating stars in the study by \cite{mathur_revisiting_2019}, among which fewer detections of acoustic modes are observed in rapidly-rotating stars. It demonstrates how stellar rotation \citep[and magnetic field, see][]{bessila_stochastic_2024} must become key parameters when predicting the observability of stellar oscillation modes \citep[see e.g.][for the PLATO mission]{goupil_predicted_2024}.

\par Using stellar modelling and oscillation numerical codes MESA and GYRE, we also find that the modes are not equally affected by rotation: the injected power is more strongly inhibited in low $n$ and low $\ell$ modes. For a given $\ell$ parameter, low-frequency modes (i.e. low $n$ modes) are more inhibited by rotation. However, when looking at the influence of $m$ or $\ell$ on the excitation, no clear tendency emerges because of the complex behaviour of the source term. (see Figures \ref{fig:puissance_modes_rotation} and \ref{fig:comparaison_m}).

\par The choice for the kinetic energy spectrum slope $\alpha$  in the theoretical prescription does not influence how rotation diminishes the power injected into the modes compared to the non-rotating case. However, the resulting power differs: the steeper the slope, the higher the power injected into the acoustic modes. Studying more precisely amplitudes of the stochastically excited acoustic modes can thus provide constraints on the turbulent kinetic energy spectrum inside the convective zone.

\par Finally, stellar structure and fundamental parameters are paramount to assessing the power injected into acoustic modes. For a fixed stellar mass and age, metallicity plays a key role. The lower-metallicity model we computed presents a higher excitation rate, nearly $10^2$ times higher than the $10$ times more metal-rich corresponding model. The main parameter when it comes to excitation is the velocity near the photosphere, where the modes are mostly excited by convection. The higher this velocity, the higher the power injected into the modes. This result needs to be further compared with observations. As highlighted in \cite{yu_asteroseismology_2018}, the effect of metallicity is difficult to disentangle from other stellar parameters such as effective temperature, mass, and luminosity. Moreover, metallicity also has an impact on rotation \citep{amard_impact_2020}: at ages superior to 1 Gyr, stars with higher metallicity are expected to spin down more efficiently. This work corroborates that in Solar-like stars, mode amplitudes are very sensitive to the mass and evolutionary stage. Therefore, to accurately assess the robustness of the present model, an extended study comparing observations to theoretical amplitudes computed using 1-D stellar models with different metallicities and rotation rates is required.

\subsection{Perspectives}
Our study opens the way to a better understanding of how acoustic modes are excited by convection, and how rotation influences the excitation. 
\par However, the mode amplitudes are a balance between driving and damping in the framework of stochastic excitation \citep{samadi_stellar_2015}. The present work focuses exclusively on the driving of such modes. Modelling the damping rate \citep[see e.g.][]{houdek_interaction_2015}, as well as the influence of rotation on the damping of the modes is key to fully assessing the impact of rotation on the mode amplitudes. To our knowledge, such a study of this phenomenon has never been done in the literature.
\par Furthermore, stellar convection zones are observed and predicted to be differentially rotating \citep[e.g.][]{thompson_internal_2003, garcia_tracking_2007, barnes_angular_2010, brun_magnetism_2017, benomar_asteroseismic_2018, brun_powering_2022}, while we considered in this study a uniform mean rotation. The effects of differential rotation on the acoustic mode excitation rates are currently under investigation (Biscarrat et al., in prep.). One can expect a different behaviour depending on the differential rotation profile (solar or anti-solar), and the differential rotation rate.
\par The present study also highlights how important the stellar structure is when it comes to mode excitation: we considered five different stellar models and witnessed changes of several orders of magnitude for the power injected into the modes. Such effect should not be ignored: in the near future, these prescriptions should be extended to a wide range of stellar models across the HR diagram. In addition, future works should also tackle PMS stars in order to select some interesting targets. Moreover, there is a strong need for better predictions of mode detection for missions such as TESS and PLATO. For example, \cite{campante_asteroseismic_2016} and \cite{goupil_predicted_2024} studied acoustic mode detection probability, taking into account the properties of the target stars. Such studies could be extended by taking rotation into account.
\par Moreover, our semi-analytical formalism can be adapted to tackle other types of waves. For gravity waves and gravito-inertial modes, the present work has to be extended to take into account the horizontal and toroidal displacements \citep{samadi_corot_2010, belkacem_solar_2019, neiner_astronomy_2020}. Such waves are stochastically excited at the base of the convective zone, which is where the Rossby number is small and the rotation effects are dominant, as demonstrated in Cartesian models by \cite{mathis_impact_2014} and \cite{augustson_model_2020}. One can then expect rotation to play an even more important role for gravity and gravito-inertial waves than for acoustic modes. One can also apply our formalism to study the impact of rotation on the stochastic excitation of Rossby waves \citep[e.g.][]{philidet_interaction_2023, blume_inertial_2024}. These waves have been recently observed in the Sun \citep{loptien_global-scale_2018, gizon_solar_2021}, and are driven by rotation. They constitute a complementary probe with acoustic modes to sound rotating stellar convection zones.
\par Finally, as shown observationally by \cite{mathur_revisiting_2019} and theoretically by \cite{bessila_stochastic_2024}, the magnetic field also has an impact on the mode excitation. Rotation strongly influences the magnetic field, which is generated by the dynamo action inside convective zones \citep[see e.g.][]{brun_magnetism_2017} and thus they are closely linked. Addressing the magnetic field effect on mode excitation is thus key to fully understanding the excitation of mode amplitudes. It should be examined first on its own and then combined with the effects of rotation.

\begin{acknowledgements}
The authors thank the referee for detailed comments that allow them to improve their work. The authors thank Jordan Philidet and Kevin Belkacem for fruitful discussions. L.B. and Stéphane M. acknowledge support from the  European  Research Council  (ERC)  under the  Horizon  Europe program (Synergy  Grant agreement 101071505: 4D-STAR), from the CNES SOHO-GOLF and PLATO grants at CEA-DAp, and from PNPS (CNRS/INSU). While partially funded by the European Union, views and opinions expressed are however those of the author only and do not necessarily reflect those of the European Union or the European Research Council. Neither the European Union nor the granting authority can be held responsible for them. Savita M.\ acknowledges support from the Spanish Ministry of Science and Innovation with the grant no. PID2019-107061GB-C66 and through AEI under the Severo Ochoa Centres of Excellence Programme 2020--2023 (CEX2019-000920-S).
\end{acknowledgements}

\bibliographystyle{aa}
\bibliography{references-2}

\appendix
\section{Rotating mixing-length theory from \cite{augustson_model_2019}}
\label{sec:appendix_mlt}

We detail here the Rotating Mixing-Length Theory (R-MLT) formalism from \cite{augustson_model_2019}, which we use in our stochastic excitation modelling to take into account the modification of the convection properties by rotation.

In this framework, we consider a local Cartesian model, which allows us to characterise the local dynamics of rotating convection around a given radius and latitude in a star or in a planet. In \cite{augustson_model_2019} the fluid dynamics is studied within a simplified Rayleigh-Bénard set-up. The key point to underline here is that scaling laws obtained within this simplified geometry have been recovered in numerical simulations of rotating stellar convection in global spherical geometry \citep{vasil_rotation_2021, korre_dynamics_2021}. Building on this, we also apply these scaling laws to model the stochastic excitation of acoustic oscillation modes in rotating solar-like pulsators.

\subsection{Linear Boussinesq equations}

\cite{augustson_model_2019} considered an infinite layer of fluid with a local rotation vector $\boldsymbol{\Omega}$, inclined with an angle $\theta$ relatively to the vertical direction. They also used the Boussinesq approximation, which neglects density fluctuations except for the buoyancy force. This means that we focus here on small-scale motions (and eddies) which have properties linked to the local conditions. The fluid has a small thermal expansion coefficient $\alpha = - (\partial \ln \rho / \partial \ln T )_P$, where $\rho$ is the density, $T$ the temperature and $P$ the pressure. The fluid is confined between two infinite impenetrable plates with different temperatures and separated by a distance $\ell_0$. In this framework, the Navier-Stokes equation within the Boussinesq approximation with rotation is written as:
\begin{equation}
    \frac{\partial \u}{\partial t}+ (\u \cdot \nabla) \u + 2 \Om \times \u = \frac{-1}{\rho} \nabla P - \boldsymbol{g} \alpha \Theta + \nu \nabla^2\u, 
\label{eq:navier_stokes}
\end{equation}
where $\u$ is the fluid velocity in the rotating frame, $p$ and $\Theta$ are the fluctuations of pressure and temperature respectively caused by convection around their hydrostatic background values. We introduce the molecular diffusivity $\nu$ and the thermal diffusivity $\kappa$.
The thermal diffusion equation is: 
\begin{equation}
    \frac{\partial \Theta}{\partial t} - \kappa \nabla^2 \Theta = \beta u_z - \u \cdot \nabla \Theta,
\end{equation}
where $\beta$ is the thermal gradient: 
\begin{equation}
    \beta = \frac{d \theta}{dz} + \frac{c_p}{g},
\end{equation}
with $c_p$ being the specific heat capacity at constant pressure.
In addition, the velocity field in the Boussinesq approximation is solenoidal: 
\begin{equation}
  \nabla \cdot \v = 0.  
  \label{eq:continuity}
\end{equation}

Linearising and combining the previous equations Eqs. (\ref{eq:navier_stokes})-(\ref{eq:continuity}), one obtains a single equation for the vertical component of the velocity \citep[see e.g.][for more details regarding its derivation]{chandrasekhar_hydrodynamic_1961}: 

\begin{equation}
    \begin{gathered}
\left(\partial_t-\kappa \nabla^2\right)\left(\partial_t-\nu \nabla^2\right)^2 \nabla^2 u_z+g \alpha_T \beta \nabla_{\perp}^2\left(\partial_t-\nu \nabla^2\right) u_z \\
+4 \Omega \cdot \nabla\left[\Omega \cdot \nabla\left(\partial_t-\kappa \nabla^2\right) u_z\right]=0.
\label{eq:dispersion_original}
\end{gathered}
\end{equation}

\noindent Furthermore, \cite{augustson_model_2019} assume impenetrable boundary velocity conditions and stress-free boundary temperature conditions, which require that the vertical wavenumber is $k_z = n \pi/\ell_0$, with $n$ an integer. We neglect in our study all diffusive processes that is $\kappa = \nu = 0$. Indeed, including such diffusive processes does not qualitatively change the results in R-MLT. The vertical velocity behaves as $\exp({st})$, where $s$ is a growth rate for the convective instability
Eq. (\ref{eq:dispersion_original}) then yields the dispersion relation that links $s$ to the wavevector $\boldsymbol{k}$:

\begin{equation}
    \hat{s}^2 + \mathcal{O}^2 \cos^2(\theta) - \frac{(z^3-1)}{z^3}   = 0,
    \label{eq:dispersion_relation_normalised}
\end{equation}
where

\begin{align}
    & N_{*}^2 = \lvert g \alpha \beta \rvert, \\
    & \hat{s} = \frac{s}{N_{*}}, \\
    & z^3 = 1 + a^2 = \frac{k^2}{k_z^2},\\
    & a^2 = \frac{k_x^2}{k_z^2} + \frac{k_y^2}{k_z^2} = a_x^2 + a_y^2, \\
    &\mathcal{O}^2 = \frac{4 \Omega^2}{N_{*}^2}.
\end{align}

Moreover, to compare the results from the R-MLT and the standard MLT as implemented in stellar evolution codes, we need to compute the characteristic convective velocity  $v_0$ without rotation. $v_0$ is derived from the growth rate and maximising wavevector in the non-rotating and non-diffusive case. This leads to $s_0^2=3 / 5\left|g_0 \alpha \beta_0\right|$, where $\beta_0$ and $g_0$ are the thermal gradient and effective gravity in the non-rotating case, respectively. From Eq. (36) in \cite{stevenson_turbulent_1979}, we have $k_0^2=5 / 2 k_z^2$ and:

\begin{equation}
    v_0=\frac{s_0}{k_0}=\frac{\sqrt{6}}{5} \frac{N_{*, 0}}{k_z}=\frac{\sqrt{6}}{5 \pi} \ell_0 N_{*, 0}.
\end{equation}

We can then define the convective Rossby number $\mathcal{R}o$:

\begin{equation}
    \mathcal{R}o=\frac{v_0}{2 \Omega_0 \ell_0 \cos \theta}=\frac{\sqrt{6} N_{*, 0}}{10 \pi \Omega_0 \cos \theta}.
\end{equation}

It implies that:

\begin{equation}
    \mathcal{O}=\frac{2 \Omega_0}{N_*}=\frac{u_0}{N_* \mathcal{R}o \ell_0}=\frac{\sqrt{6} N_{*, 0} \cos \theta}{5 \pi N_* \mathcal{R}o}.
\end{equation}

We then consider the variation of the superadiabaticity, which is given by $\epsilon=H_P \beta / T$. With this definition, $N_*^2=\left|g \alpha T \epsilon / H_P\right|$, where $H_P$ is the pressure scale height. We then introduce $N_{*, 0} = \left|g_0 \alpha \beta_0\right|$ in the non-rotating case. So far, all quantities have been normalised with respect to $N_*$, which includes the impact of rotation on $\beta$ and $g$. We introduce the ratio of superadiabaticities as an additional unknown:

\begin{equation}
    q=N_{*, 0} / N_*.
\end{equation}

\noindent Therefore, one has:
\begin{equation}
     \mathcal{O}=q \frac{\sqrt{6}}{5 \pi \mathcal{R}o}=q \cos \theta \mathcal{O}_0,
\end{equation}
where: 

\begin{equation}
    \mathcal{O}_0 = \frac{ 4 \Omega^2}{N_{*,0}^2}.
\end{equation}

\subsection{Heat-flux maximisation}

The method defined by \cite{malkus_heat_1954} relies on the heat-flux maximisation principle: it assumes that the linear mode that maximises the total convective heat flux is dominant. With Mixing-Length Theory, it amounts to selecting this maximising mode to compute the characteristic convective velocity and wavenumber for the flow. In the absence of any diffusive process, the convective heat-flux writes \citep[see e.g.][]{stevenson_turbulent_1979}: 

\begin{equation}
\mathcal{F}=\frac{\mathcal{F}_0}{q^3}\frac{\hat{s}^3}{z^3},
\label{eq:def_flux}
\end{equation}

\noindent where $\mathcal{F}_0=\langle\rho\rangle c_P N_{*, 0}^3 /\left(g \alpha k_z^2\right)$ is the heat-flux in the absence of rotation. When maximised for a positive real value of $\hat{s}$, Eq. (\ref{eq:def_flux}) implies that
\begin{equation}
    \frac{d \hat{s}}{d z}=\frac{\hat{s}}{z}.
    \label{eq:heat_max}
\end{equation}

The derivative of the dispersion relationship yields another expression for $d \hat{s} / d z$:

\begin{equation}
   \frac{d\hat{s}}{dz} = \frac{- 3z^2 \hat{s}(\hat{s}^2-1)}{1 + \mathcal{O}_o^2 q^2  + z^3(3 \hat{s}^2 -1)}.
   \label{eq:dispersion}
\end{equation}

\noindent Moreover, \cite{augustson_model_2019} used a heat flux invariance approximation. Indeed, the total heat flux is assumed to remain identical with and without rotation (see section 2.3 from \cite{augustson_model_2019} for more details). This condition yields:

\begin{equation}
\frac{\max [\mathcal{F}]}{\max [\mathcal{F}]_0} = \frac{25}{6} \sqrt{\frac{5}{3}} \frac{\hat{s}^3}{q^3 z^3}=1,
\end{equation}
which implies:
\begin{equation}
    \hat{s}=\tilde{s} q z,
\label{eq:heat_invariance}
\end{equation}
where $\tilde{s}=2^{1 / 3} 3^{1 / 2} 5^{-5 / 6}$.
One then has the convective velocity and wavenumber modulation, depending only on the maximising value for $z$: 
\begin{equation}
    \frac{k}{k_0} = \frac{k_z z^{3/2}}{k_0} = \sqrt{2/5} z^{3/2},
    \label{eq:k}
\end{equation}
\begin{equation}
    \frac{u}{u_0}=\frac{k_0}{s_0} \frac{s}{k}=\left(\frac{5}{2}\right)^{\frac{1}{6}} z^{-\frac{1}{2}}.
    \label{eq:v}
\end{equation}

\noindent Finally, only the maximising wavenumber $z$ is needed. Equating the right-hand sides of Eqs. (\ref{eq:heat_max}) and (\ref{eq:dispersion}) then applying (\ref{eq:heat_invariance}), we find:

\begin{equation}
    2 z^5 - 5 z^2 - \frac{18}{25 \pi^2 \mathcal{R}o^2 \tilde{s}^2} = 0,
\end{equation}
which is Eq.(46) from \cite{augustson_model_2019}. Together with Eqs. (\ref{eq:k}) and (\ref{eq:v}), one can recover the convective velocity (resp. convective wavenumber) modified by rotation in the framework of R-MLT. 

\section{MESA inlists for the solar model}
\label{sec:inlist}
\begin{verbatim}
&kap
  ! kap options
  ! see kap/defaults/kap.defaults
  use_Type2_opacities = .true.
    Zbase = 0.016

/ ! end of kap namelist

&controls

      initial_mass = 1.0 

      ! MAIN PARAMS
      mixing_length_alpha = 1.9446893445
      initial_z = 0.02 ! 0.04, 0.002, 0.026
      do_conv_premix = .true.
      use_Ledoux_criterion = .true.

      ! OUTPUT
      max_num_profile_models = 100000
      profile_interval = 300
      history_interval = 1
      photo_interval = 300

      ! WHEN TO STOP
      xa_central_lower_limit_species(1) = 'h1'
      xa_central_lower_limit(1) = 0.01
      max_age = 6.408d9

      ! RESOLUTION
      mesh_delta_coeff = 0.5
      time_delta_coeff = 1.0

      ! GOLD TOLERANCES
      use_gold_tolerances = .true.
      use_gold2_tolerances = .true.
      delta_lg_XH_cntr_limit = 0.01
      min_timestep_limit = 1d-1

      !limit on magnitude
      delta_lgTeff_limit = 0.25 ! 0.005
      delta_lgTeff_hard_limit = 0.25 ! 0.005
      delta_lgL_limit = 0.25 ! 0.005

      ! asteroseismology
      write_pulse_data_with_profile = .true.
      pulse_data_format = 'FGONG'
      ! add_atmosphere_to_pulse_data = .true.


/ ! end of controls namelist


&pgstar



/ ! end of pgstar namelist
\end{verbatim}

\clearpage
\section{Hertzsprung-Russell diagram and profiles for the MESA models}

\begin{figure}[h!]
    \centering
    \includegraphics[width = \linewidth]{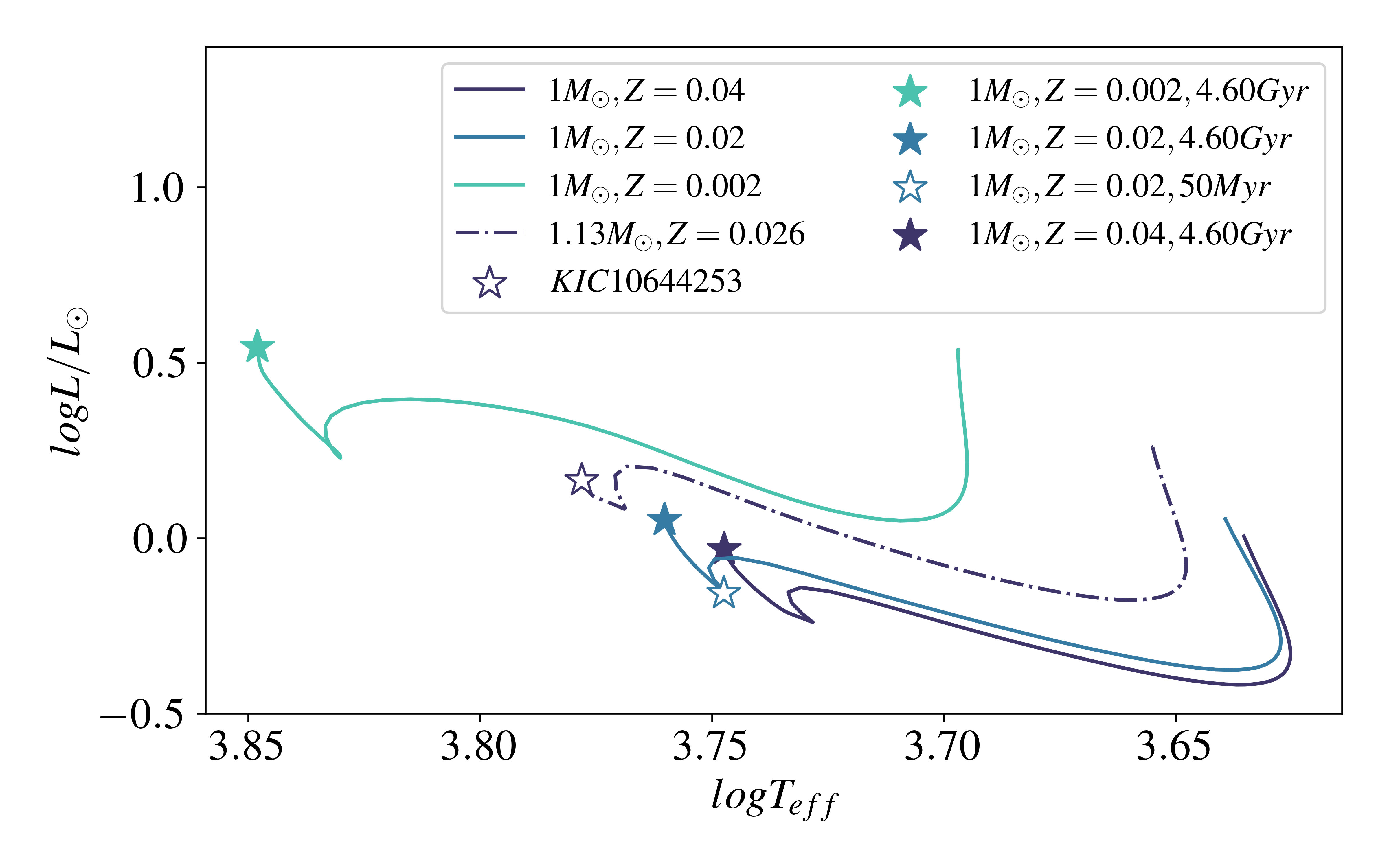}
    \caption{Hertzsprung–Russell diagram for the computed MESA models }
    \label{fig:hr_diagram}
\end{figure}

\begin{figure}[h!]
    \centering
    \includegraphics[width = \linewidth]{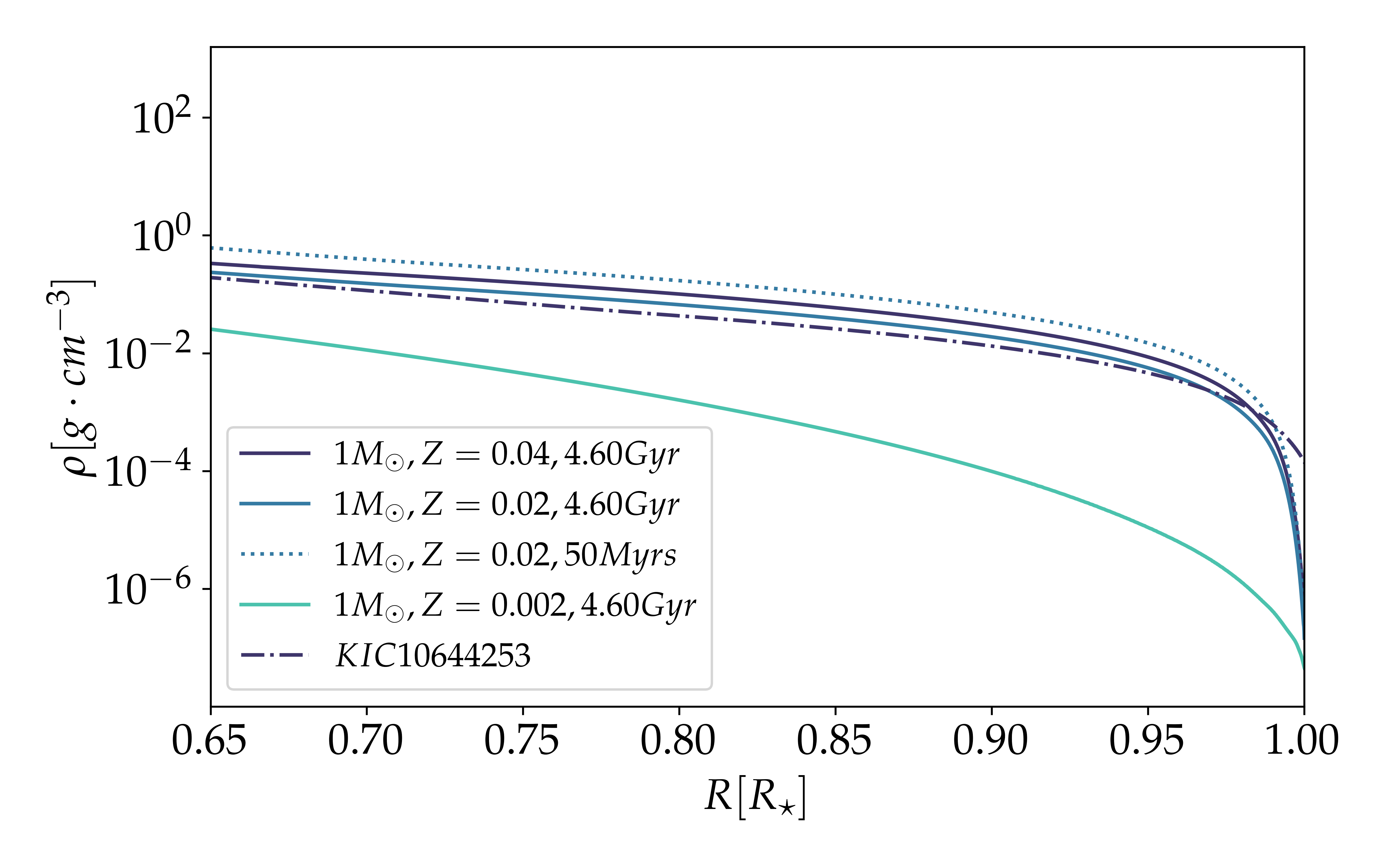}
    \includegraphics[width = \linewidth]{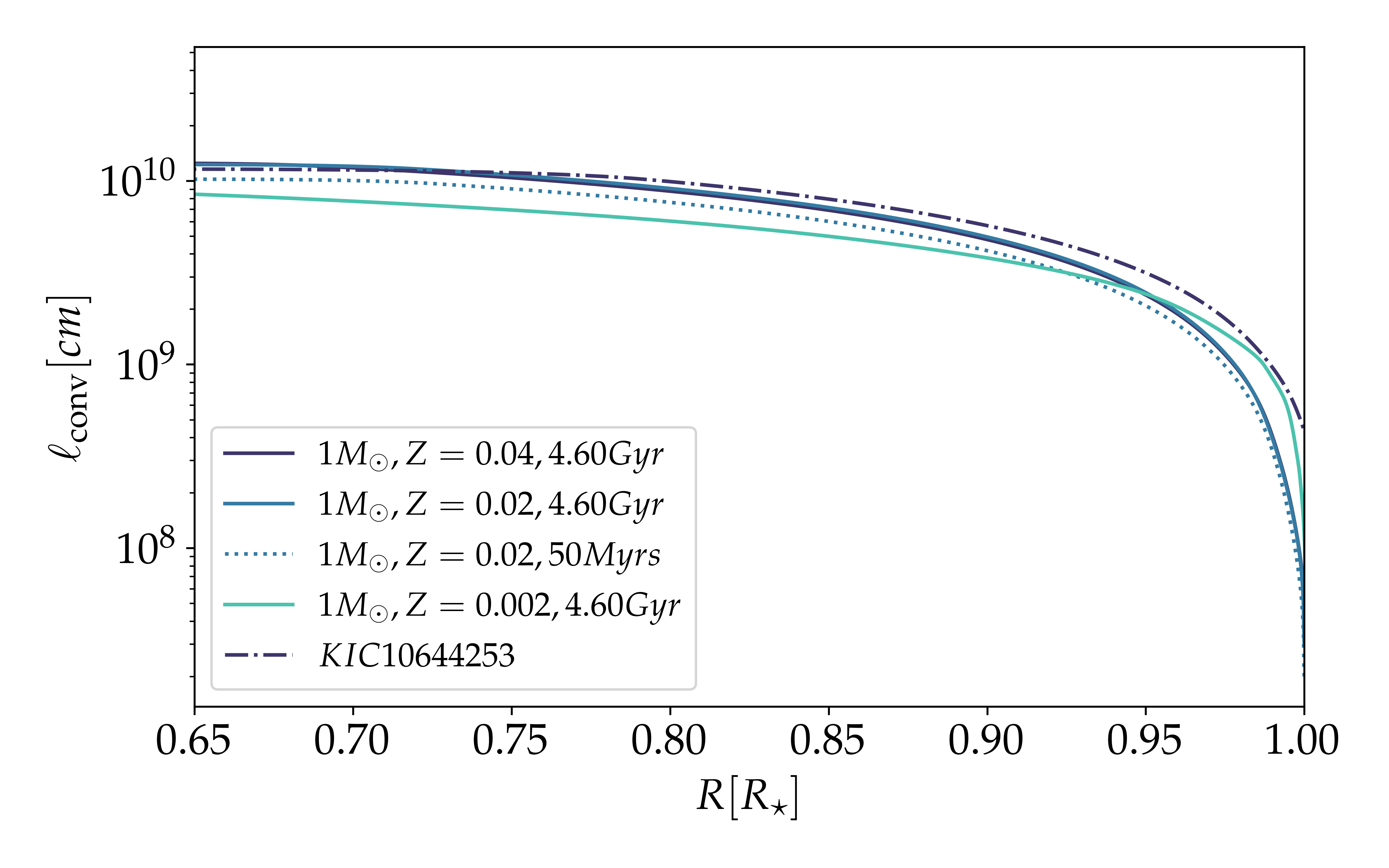}
    \includegraphics[width =\linewidth]{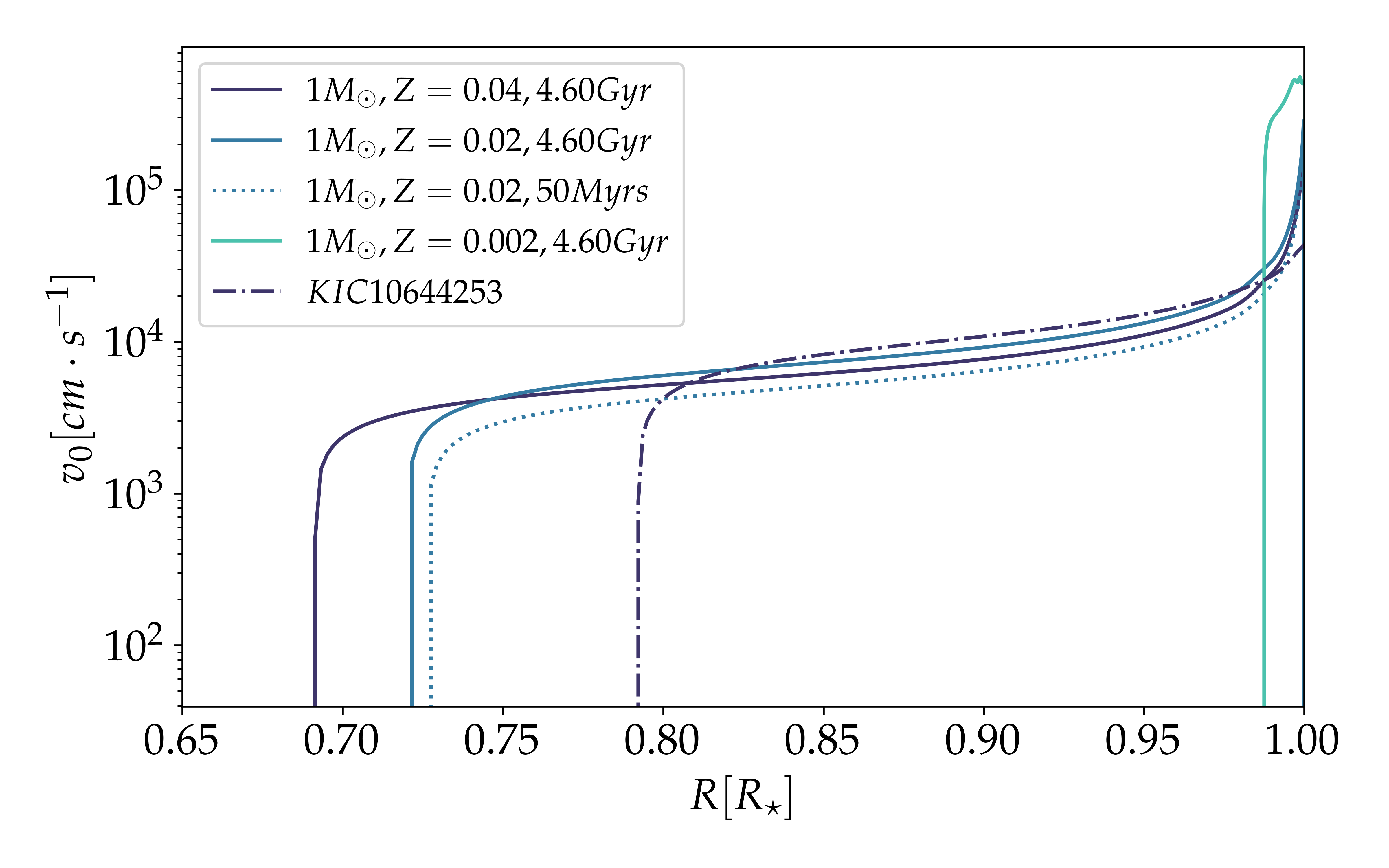}
    \caption{(Top) Radial profiles for the density. (Centre) Radial profile for the convective mixing-length profile. (Bottom) Radial profile for the convective velocity.}
    \label{fig:profiles}
\end{figure}

We show here the Hertzsprung-Russell diagram for the MESA models we computed in Fig. \ref{fig:hr_diagram}, as well as the profiles for the density, Mixing-Length from non-rotating MLT and convective velocity from non-rotating MLT in Fig. \ref{fig:profiles}. The stellar properties for these models are detailed in Tab. \ref{tab:stellar_properties}.

\begin{table}[h!]
\setlength{\tabcolsep}{1.6pt} 
\renewcommand{\arraystretch}{1.5} 
\centering
\begin{tabular}{c|c|c|c|c|c}
Mass  & Age & Z  & $T_{\rm eff}$  & Radius  & Convective boundary \\ 
($M_{\odot}$) &  & (dex) & (K) & ($R_{\odot}$) & ($R_{CZ}/R_{\star}$) \\ \hline \hline
1 & 4.6 Gyrs & 0.002 & 7048 & 1.26 & 0.99 \\ 
1 & 4.6 Gyrs & 0.02 & 5758 & 1.01 & 0.72 \\ 
1 & 4.6 Gyrs & 0.04 & 5589 & 1.06 & 0.69 \\ 
1 & 50 Myrs & 0.02 & 5590 & 0.89 & 0.73 \\ 
1.13 & 1.32 Gyrs & 0.026 & 6005 & 1.120 & 0.79 \\ 
\end{tabular}
\caption{Table of Stellar Properties for the computed MESA models.}
\label{tab:stellar_properties}
\end{table}

\nolinenumbers
\end{document}